\documentclass[preprint2]{aastex61}
\usepackage{hyperref}
\usepackage{amsmath}
\usepackage[flushleft]{threeparttable}
\usepackage{color} 

\shorttitle{CARMA-NRO Orion Survey}
\shortauthors{Kong et al.}

\begin{document}

\title{The CARMA-NRO Orion Survey}

\author{Shuo Kong}
\affiliation{Department of Astronomy, Yale University, New Haven, Connecticut 06511, USA}

\author{H\'ector G. Arce}
\affiliation{Department of Astronomy, Yale University, New Haven, Connecticut 06511, USA}

\author{Jesse R. Feddersen}
\affiliation{Department of Astronomy, Yale University, New Haven, Connecticut 06511, USA}

\author{John M. Carpenter}
\affiliation{Joint ALMA Observatory, Alonso de C\'ordova 3107 Vitacura, Santiago, Chile}

\author{Fumitaka Nakamura}
\affiliation{National Astronomical Observatory of Japan, 2-21-1 Osawa, Mitaka, Tokyo 181-8588, Japan}

\author{Yoshito Shimajiri}
\affiliation{Laboratoire AIM, CEA/DSM-CNRS-Universit$\acute{\rm e}$ Paris Diderot, IRFU/Service d'Astrophysique, CEA Saclay, F-91191 Gif-sur-Yvette, France}

\author{Andrea Isella}
\affiliation{Department of Physics and Astronomy, Rice University, 6100 Main Street, Houston, TX, 77005, USA}

\author{Volker Ossenkopf-Okada}
\affiliation{I.~Physikalisches Institut, Universit\"at zu K\"oln,
              Z\"ulpicher Str. 77, D-50937 K\"oln, Germany}

\author{Anneila I. Sargent}
\affiliation{California Institute of Technology, Cahill Center for Astronomy and Astrophysics 249-17, Pasadena, CA 91125, USA}

\author{\'Alvaro S\'anchez-Monge}
\affiliation{I.~Physikalisches Institut, Universit\"at zu K\"oln,
              Z\"ulpicher Str. 77, D-50937 K\"oln, Germany}

\author{S\"umeyye T. Suri}
\affiliation{I.~Physikalisches Institut, Universit\"at zu K\"oln,
              Z\"ulpicher Str. 77, D-50937 K\"oln, Germany}

\author{Jens Kauffmann}
\affiliation{Haystack Observatory, Massachusetts Institute of Technology, 99 Millstone Road, Westford, MA 01886, USA}

\author{Thushara Pillai}
\affiliation{Max--Planck--Institut f\"ur Radioastronomie, Auf dem H\"ugel 69, D--53121 Bonn, Germany}

\author{Jaime E. Pineda}
\affiliation{Max-Planck-Institut f\"ur extraterrestrische Physik, Giessenbachstrasse 1, 85748 Garching, Germany}

\author{Jin Koda}
\affiliation{Department of Physics and Astronomy, Stony Brook University, Stony Brook, NY 11794-3800}

\author{John Bally}
\affiliation{Department of Astrophysical and Planetary Sciences, University of Colorado, Boulder, Colorado, USA}

\author{Dariusz C. Lis}
\affiliation{Sorbonne Universit\'{e}, Observatoire de Paris, Universit\'{e} PSL, CNRS, LERMA, F-75014, Paris, France}
\affiliation{California Institute of Technology, Cahill Center for Astronomy and Astrophysics 301-17, Pasadena, CA 91125, USA}

\author{Paolo Padoan}
\affiliation{Institut de Ci\`{e}ncies del Cosmos, Universitat de Barcelona, IEEC-UB, Mart\'{i} i Franqu\`{e}s 1, E08028 Barcelona, Spain; ppadoan@icc.ub.edu}
\affiliation{ICREA, Pg. Llu\'{i}s Companys 23, 08010 Barcelona, Spain}

\author{Ralf Klessen}
\affiliation{Universit\"{a}t Heidelberg, Zentrum f\"{u}r Astronomie, Albert-Ueberle-Str. 2, 69120 Heidelberg, Germany}
\affiliation{Universit\"{a}t Heidelberg, Interdisziplin\"{a}res Zentrum f\"{u}r Wissenschaftliches Rechnen, INF 205, 69120 Heidelberg, Germany}

\author{Steve Mairs}
\affiliation{East Asian Observatory, 660 N. A'ohoku Place, Hilo, Hwaii, 96720, s.mairs@eaobservatory.org}

\author{Alyssa Goodman}
\affiliation{Harvard-Smithsonian Center for Astrophysics, 60 GARDEN STREET, MS 42, CAMBRIDGE, MA 02138, USA}

\author{Paul Goldsmith}
\affiliation{Jet Propulsion Laboratory, California Institute of Technology, 4800 Oak Grove Drive, Pasadena, CA 91109, USA}

\author{Peregrine McGehee}
\affiliation{Department of Earth, Space, and Environment Sciences, College of the Canyons, Santa Clarita, CA 91355}

\author{Peter Schilke}
\affiliation{I.~Physikalisches Institut, Universit\"at zu K\"oln,
              Z\"ulpicher Str. 77, D-50937 K\"oln, Germany}

\author{Peter J. Teuben}
\affiliation{Astronomy Department, University of Maryland, College Park, MD 20742, USA}

\author{Mar\'ia Jos\'e Maureira}
\affiliation{Department of Astronomy, Yale University, New Haven, Connecticut 06511, USA}

\author{Chihomi Hara}
\affiliation{Department of Astronomy, The University of Tokyo, 7-3-1 Hongo Bunkyo, Tokyo 113-0033, Japan}

\author{Adam Ginsburg}
\affiliation{National Radio Astronomy Observatory,  1003 Lopezville road, Socorro NM, 87801}

\author{Blakesley Burkhart}
\affiliation{Institute for Theory and Computation, Harvard-Smithsonian Center for Astrophysics, 60 Garden St., Perkin Lab 240, Cambridge, MA 02138}

\author{Rowan J. Smith}
\affiliation{Jodrell Bank Centre for Astrophysics, School of Physics and Astronomy, University of Manchester, Oxford Road, Manchester M13 9PL, UK}

\author{Anika Schmiedeke}
\affiliation{Max-Planck-Institut f\"ur extraterrestrische Physik, Giessenbachstrasse 1, 85748 Garching, Germany}
\affiliation{I.~Physikalisches Institut, Universit\"at zu K\"oln,
              Z\"ulpicher Str. 77, D-50937 K\"oln, Germany}

\author{Jorge L. Pineda}
\affiliation{Jet Propulsion Laboratory, California Institute of Technology, 4800 Oak Grove Drive, Pasadena, CA 91109, USA}

\author{Shun Ishii}
\affiliation{Joint ALMA Observatory, Alonso de C\'ordova 3107 Vitacura, Santiago, Chile}

\author{Kazushige Sasaki}
\affiliation{Department of Physics, Niigata University, 8050 Ikarashi-2, Niigata 950-2181, Japan}

\author{Ryohei Kawabe}
\affiliation{National Astronomical Observatory of Japan, 2-21-1 Osawa, Mitaka, Tokyo 181-8588, Japan}

\author{Yumiko Urasawa}
\affiliation{Department of Physics, Niigata University, 8050 Ikarashi-2, Niigata 950-2181, Japan}

\author{Shuri Oyamada}
\affiliation{Japan Women’s University, 2-8-1 Mejirodai, Bunkyo-ku, Tokyo
112-8681, Japan}

\author{Yoshihiro Tanabe}
\affiliation{College of Science, Ibaraki University, 2-1-1 Bunkyo, Mito, Ibaraki 310-8512, Japan}

\begin{abstract}

We present the first results from a new, high resolution, 
$^{12}$CO(1-0), $^{13}$CO(1-0), and C$^{18}$O(1-0)
molecular line survey of the Orion A cloud, 
hereafter referred to as the CARMA-NRO Orion Survey. 
CARMA observations have been combined with single-dish 
data from the Nobeyama 45m telescope to provide 
extended images at about 0.01 pc resolution, 
with a dynamic range of approximately 1200 in spatial scale.
Here we describe the practical details of the data 
combination in uv space, including flux scale matching, 
the conversion of single dish data to visibilities, and
joint deconvolution of single dish and 
interferometric data. A $\Delta$-variance
analysis indicates that no artifacts are
caused by combining data from the two instruments. 
Initial analysis of the data cubes, including 
moment maps, average spectra, channel maps,
position-velocity diagrams, excitation temperature, 
column density, and line ratio maps
provides evidence of complex 
and interesting structures such as filaments, bipolar 
outflows, shells, bubbles, and photo-eroded pillars. 
The implications for star formation processes 
are profound and follow-up scientific studies 
by the CARMA-NRO Orion team are now underway. 
We plan to make all the data products described here 
generally accessible; some are already available at 
\dataset[https://dataverse.harvard.edu/dataverse/CARMA-NRO-Orion]{https://dataverse.harvard.edu/dataverse/CARMA-NRO-Orion}.

\end{abstract}


\section{Introduction}

Star formation (SF) in the Milky Way 
takes place in giant molecular 
clouds (GMCs) that can contain 
10$^5$ $\rm M_\odot$ of gas. 
As described in a recent review by  
\citet{2015ARA&A..53..583H}, 
a complete understanding of the properties 
of these clouds and their overall role 
in star formation has been a major 
goal for over 50 years. Nevertheless, 
many questions remain. A key issue is 
rooted in the high spatial dynamic range 
of star formation processes, where the 
span of physical scales runs from 
$\sim$ 10 pc for GMCs to $\sim$ AU for 
proto-stars. This dramatic contrast, a 
factor of $\sim$ 10$^6$, demands high dynamic
range observations. By combining molecular 
line interferometer and single-dish observations 
it is now possible to map extended areas of 
molecular clouds and simultaneously probe 
small-scale SF scenarios within the 
larger-scale structures of the GMC gas.
The resulting high dynamic range images can 
provide the comprehensive picture of star 
formation in nearby GMCs that is critical 
to improving our understanding of the many 
processes involved in the formation of stars.
Topics that can be more fully addressed include 
the roles of turbulence and magnetic fields in SF, 
the overall efficiency of the SF process 
and the effects of feedback mechanisms. 

For example, it is well known that GMCs are 
highly turbulent \citep[e.g.,][]{2004RvMP...76..125M,2007ARA&A..45..565M,2014prpl.conf...77P,2016SAAS...43...85K}.
A variety of explanations for the observed filamentary 
structures \citep{2018MNRAS.473.4220L} and velocity fields 
\citep{2012A&ARv..20...55H}, as well as signatures 
such as the mass surface density Probability Distribution
Function 
\citep[$\Sigma$-PDF,][]{2015MNRAS.448.3297F} 
have been put forward, based on numerical simulations
\citep[also see][]{2010A&A...512A..81F}. 
Large-scale mapping of the molecular gas enables 
studies of turbulence on all scales and provides 
statistically meaningful results for comparison 
with these simulations. Other signatures, such as 
the effect of magnetic fields on large-scale gas 
kinematics, may also be examined. Most importantly, 
the high dynamic range images provide much-needed 
information for GMCs with active SF across the 
whole mass spectrum.

Likewise, large-scale molecular gas maps encompassing 
a variety of cloud environments make it possible 
to study the formation and evolution of young stars 
in detail, from the first stages of dense core 
formation through the various proto-stellar phases. 
In fact, at any given time only a few percent of 
the cloud mass is of sufficiently high density to 
form stars \citep{2012ARA&A..50..531K}. Most of 
the mass resides in low density quiescent areas
\citep[see, e.g.,][]{2005ApJ...630..250K}. 
Understanding how the cloud material transitions 
from a low- to high density state, and why such 
a small fraction of the mass is confined to dense 
cores, is critical to determining the physical 
processes that control the star formation rate 
in molecular clouds. Numerical simulations suggest 
that turbulence, magnetic fields, and feedback 
effects can all cause inefficient SF
\citep[e.g.,][]{2007ApJ...662..395N,2009ApJ...693..250B,2012ApJ...750...13C,2012ApJ...759L..27P,2014prpl.conf..243K,2015MNRAS.450.4035F,2017ApJ...840...48P,2018MNRAS.476..771C,2017NatAs...1E.158L}. 
Detailed large-scale images of various molecular 
tracers can complement these simulations. 
In particular, CO isotopologue maps can inform 
studies of feedback effects such as stellar winds 
\citep{2011ApJ...742..105A,2015ApJS..219...20L},
outflows \citep{2010ApJ...715.1170A,2011MNRAS.418.2121G,2012MNRAS.425.2641N},
radiative heating \citep{2015ApJ...812...75G}, 
and supernova \citep{2017A&A...604A..13C}.

The Orion A cloud provides an ideal laboratory to 
map an extended molecular cloud in detail and 
address many of the scientific questions raised above. 
An active, star-forming region where both high- 
and low-mass young stellar 
objects (YSOs) are observed, the Orion A cloud, 
encompasses a range of star-forming environments 
\citep{2008hsf1.book..459B}. It is a huge mass reservoir
\citep[10$^5$ $\rm M_\odot$;][]{2008hsf1.book..459B} 
displaying a range 
of kinematic and dynamic characteristics, 
numerous features of feedback effects including 
shocks, jets, outflows, bubbles, and 
photon-dominated regions (PDRs). Perhaps most 
importantly,  at a distance of about 400 pc it 
is the closest site of massive star formation. 
As a result, high spatial dynamic range imaging 
of the molecular gas in Orion A will enable 
detailed comparison with a wide variety of numerical 
simulations covering physics and chemistry, 
low-mass SF to high-mass SF, and isolated and 
binary SF as well as cluster formation.

Here we present the CARMA-NRO Orion Survey, 
our large-scale, high dynamic-range imaging 
program for CO isotopologues in the 
Orion A cloud. To date there have been a significant number of 
large scale surveys of the Orion cloud at millimeter and far infrared wavelengths. 
These, and their corresponding spatial resolutions, are listed in Table 1 for comparison purposes.
Using the Combined Array for Research in Millimeter Astronomy 
(CARMA), we have mapped a 2 deg$^2$ 
region of the cloud in $^{12}$CO J=1-0, $^{13}$CO J=1-0, 
C$^{18}$O J=1-0, CN 1-0,J=3/2-1/2, SO 2(3)-1(2), 
CS J=2-1, and 3 mm continuum emission. The results for millimeter dust continuum and other molecular species
will be presented in a future paper. The same region was mapped in
$^{12}$CO J=1-0, $^{13}$CO J=1-0, C$^{18}$O J=1-0 
using the Nobeyama Radio Observatory 45 m 
telescope (NRO45). Our broad goal was
to combine the CARMA and NRO45 images and
trace the gas kinematics on scales from 
$\sim 16$ pc down to about 0.01 pc, corresponding 
to $\sim 2000$ AU at our adopted distance of 400 pc 
\footnote{Recent estimates of the distance to the Orion Nebula
Cluster (ONC) lie between 414 pc \citep{2007A&A...474..515M} 
and 388 pc \citep{2017ApJ...834..142K}.}. 
The resulting images are the first molecular line maps 
of a GMC at such large spatial dynamic range ($\sim$ 1200) 
and with a resolution that is sufficient to study even 
protostellar envelopes in detail. 

\citet{1987ApJ...312L..45B} mapped the Orion A cloud in
$^{13}$CO J=1-0 with the Bell Lab 7m telescope. 
Their map, at the resolution of 97\arcsec,
covers a larger region than our map, including
the integral-shaped filament 
and the entire L1641 cloud (i.e., from 
about DEC = -4\arcdeg to DEC = -9\arcdeg).
\citet{Ripple2013} mapped the Orion A cloud in
$^{12}$CO J=1-0 and $^{13}$CO J=1-0 with the 
Five College Radio Astronomy Observatory 
(FCRAO) 14m telescope, at the resolution of 46\arcsec. 
These maps of Orion A have significantly
lower resolution (by a factor of 5 to 10)
compared to the maps presented here.
Table \ref{tab:survey} lists these two and other
selected large-scale maps and surveys of the Orion A 
cloud observed at different wavelengths.

\begin{deluxetable*}{cccc}
\tablecolumns{4}
\tablewidth{0pt}
\tablecaption{Selected Multi-wavelength Surveys in the Orion A cloud\label{tab:survey}}
\tablehead{
\colhead{Telescope/Survey} &
\colhead{Data Type} &
\colhead{Resolution} &
\colhead{Key Reference}}
\startdata
Bell Lab 7m&$^{13}$CO(1-0)&97\arcsec &\citet{1987ApJ...312L..45B}\\  
JCMT&450$\mu m$, 850$\mu m$&7.5\arcsec, 14\arcsec &\citet{1999ApJ...510L..49J}\\
Harvard-CfA 1.2m&$^{12}$CO(1-0)&8$\farcm$4&\citet{Wilson2005}\\    
NRO 45m&H$^{13}$CO$^+$(1-0)&21\arcsec &\citet{Ikeda2007}\\
ASTE, NRO 45m&1.1 mm, $^{12}$CO(1-0)&36\arcsec, 21\arcsec &\citet{2011PASJ...63..105S}\\
Spitzer/Spitzer Orion&MIR 3-24$\mu m$&2-5\arcsec &\citet{Megeath2012}\\
JCMT/GBS&$^{13}$CO(3-2), C$^{18}$O(3-2)&17\arcsec &\citet{Buckle2012}\\
FCRAO 14m&$^{12}$CO(1-0), $^{13}$CO(1-0)&46\arcsec &\citet{Ripple2013}\\
Herschel/HOPS&FIR 70-160$\mu m$&5-12\arcsec &\citet{Fischer2013}\\
Herschel/HGBS&FIR 70-500$\mu m$&10-37\arcsec &\citet{Roy2013}\\
IRAM 30m&$^{12}$CO(2-1), $^{13}$CO(2-1)&11\arcsec &\citet{Berne2014}\\
Herschel-Planck dust&NIR, FIR, mm&2\arcsec-5\arcmin &\citet{Lombardi2014}\\
JCMT/GBS&450$\mu m$, 850$\mu m$&8\arcsec, 13\arcsec &\citet{Salji2015}\\
VISTA/VISION&NIR 0.85-2.4$\mu m$&0.8\arcsec &\citet{Meingast2016}\\
APO 2.5m/IN-SYNC&NIR 1.5-1.6$\mu m$&1.6\arcsec &\citet{DaRio2016}\\
GBT/GAS&NH$_3$(1,1) (2,2) (3,3)&32\arcsec &\citet{KirK2017}\\
JCMT/BISTRO&850$\mu m$ polarization&14\arcsec &\citet{Ward-Thompson2017}\\
CARMA+NRO 45m/CARMA-NRO Orion&$^{12}$CO(1-0), $^{13}$CO(1-0), C$^{18}$O(1-0)&8\arcsec &this paper\\
\enddata
\tablecomments{The surveys included cover at least the integral-shaped filament \citep{1987ApJ...312L..45B}. In the first column, the survey name, when it exists, is included to the right of the telescope used for the survey.}
\end{deluxetable*}
\section{Observations}\label{sec:obs}

\subsection{CARMA}\label{subsec:carmaobs}

The CARMA 23-element millimeter wavelength 
array comprised six 10-m diameter
antennas, nine 6-m antennas, and eight 3.5-m antennas. 
For this survey, observations were obtained 
with the 10-m and 6-m antennas in the CARMA ``D'' and
``E'' configurations to provide uv coverage 
encompassing 7-104m (2.5-40k$\lambda$ where
$\lambda$ = 2.6 mm at 115 GHz),
corresponding to angular scales of 5\arcsec-70\arcsec. 

The region surveyed encompasses an area of approximately
2 deg$^2$ toward the Orion A cloud, 
bounded by the OMC-3 in the north, 
and L1641-C in the south.
This was divided into 181 subfields, 
each of 6\arcmin$\times$6\arcmin\ size. 
Each subfield was covered in 126 pointings 
observed with a Nyquist sampling in an 
hexagonal grid, for a total of 22,806
pointings over 650 hours.
For one pass through each 
subfield the integration time was
typically 15-20 seconds per pointing. 
For each array configuration most
subfields were observed three times at 
hour angles spaced by 2 h to maximize the
$u,v$ coverage. Hour angle 
coverage in the two configurations was
shifted by 1 h, such that most subfields were
observed 6 times at 6 different hour angles.

Observations were carried out  
between 2013 and 2015 as a CARMA Key
Project. The receiver was tuned to 
local oscillator (LO) frequency of 109.41
GHz and the correlator was configured to 
cover the $^{12}$CO J=1-0, $^{13}$CO
J=1-0, C$^{18}$O J=1-0, CN 1-0,J=3/2-1/2, 
and SO 2(3)-1(2) lines in the upper sideband (USB) and
CS J=2-1 in the lower side band (LSB). 
The correlator windows for $^{12}$CO
J=1-0 and CN used  
a bandwidth of 31 MHz (81 $\rm
km~s^{-1}$) with spectral resolution 
of 98 kHz ($\sim$ 0.25 $\rm km~s^{-1}$).
The correlator windows for the remaining 
spectral lines had bandwidths of 8
MHz ($\sim 21$ $\rm km~s^{-1}$) and
spectral resolution of 24 kHz ($\sim 0.067$
$\rm km~s^{-1}$). The correlator setup also 
included two 500 MHz bandwidth windows 
centered at 110.8 GHz and 111.3 GHz in USB,
with the image sideband centered at 103.0 GHz and 
102.5 GHz, for a total of 2 GHz continuum bandwidth.
In this paper, only the $^{12}$CO J=1-0, $^{13}$CO J=1-0, 
and C$^{18}$O J=1-0 CARMA data are presented. 

Calibration of the bandpass for each spectral window relied primarily on observations
of 3C84, but the quasars 0423-013 and 0532+075 were also used. 
Uranus was the primary flux
calibrator. When it was not available, 
Mars or observations of 3C84 calibrated from
prior observations were used. Gain calibration 
relied on observations of 0532+075 every
$\sim$ 20 minutes. 
The MIRIAD \citep{1995ASPC...77..433S} 
task {\tt mossdi} was used to create maps from
the CARMA observations (with a Briggs weighting of 0.5) with an angular 
resolution of $\sim$ 7\arcsec\ and sensitivity of 0.62 Jy/beam 
in a 0.25 $\rm km~s^{-1}$ channel for $^{12}$CO,
and 0.50 Jy/beam in a 0.10 $\rm km~s^{-1}$ 
channel for $^{13}$CO and C$^{18}$O.

\subsection{NRO45}

NRO45 mapping observations of the Orion A cloud 
in the $^{12}$CO(1-0), $^{13}$CO(1-0), 
and C$^{18}$O(1-0) lines were made using two 
receivers: 1) BEARS, a 25-beam receiver; and 2) FOREST, 
a newer 4-beam dual polarization, 
sideband separating, SIS receiver 
\citep{2016SPIE.9914E..1ZM}. The telescope 
beam size is $\sim$15\arcsec\ at 115 GHz 
and the main beam efficiency $\eta_{\rm MB}$ 
is $\sim$27--43\%.

BEARS observations of the Orion A cloud
in $^{12}$CO(1-0) were acquired between 
December 2007 and January 2009. 
The isotopologues $^{13}$CO(1-0) and 
C$^{18}$O(1-0) were observed between 
May 2013 and January 2014. Some of these 
data have already been published \citep{2011PASJ...63..105S,2012ApJ...746...25N,2014A&A...564A..68S,2015ApJS..217....7S}.
A combination of 25 sets of 1024 channel 
auto-correlators (one for each receiver) with 32 MHz bandwidth 
provided frequency resolution of 31.25 kHz 
\citep{2000SPIE.4015...86S}, corresponding 
to $\sim$0.1 km s$^{-1}$ velocity resolution 
at 115 GHz. Standard on-the-fly (OTF) 
mapping techniques, with off-position 
(RA$_{\rm J2000}$, DEC$_{\rm J2000}$) = 
($\rm 5^h29^m00\fs0,-5\arcdeg25\arcmin30\farcs0$)
were used to produce images of the extended cloud. 
To minimize any adverse effects due to the process, 
separate OTF maps were obtained by scanning in the
RA and DEC directions.
Pointing accuracy was 
better than 3$\arcsec$ throughout the observations.  

NRO45 mapping observations of the Orion A 
cloud using the FOREST receiver were carried out between 
December 2014 and February 2017. Until December 2016, 
the $^{12}$CO(1-0), $^{13}$CO(1-0), C$^{18}$O(1-0)
molecular lines were observed simultaneously. 
Thereafter, due to a change in the receiver setting, 
only $^{13}$CO(1-0) and C$^{18}$O(1-0) were included.
 
The SAM45 spectrometer provided a bandwidth of 63 MHz and 
frequency resolution of 15.26 kHz, corresponding to 
velocity resolution $\sim$0.04 km s$^{-1}$. 
The line intensity scale was derived from direct 
comparison with the $\rm ^{12}$CO(1-0) 
and $\rm ^{13}$CO(1-0) $T_{\rm mb}$ maps obtained 
by the BEARS receiver.

During the FOREST observations, 
we mapped a small area of the OMC-2/FIR4 
region \citep[the strongest dust continuum 
emission in the OMC-2/3 region,][]{2008ApJ...683..255S,2015ApJS..217....7S,2015ApJS..221...31S} 
two or three times per day. 
We scaled the 
FOREST ($T_{\rm a}^*$) to the BEARS 
($T_{\rm mb}$) by comparing the FOREST 
intensity with the BEARS intensity in 
the OMC-2/FIR 4 region.
As for the BEARS observations, the FOREST observations 
were obtained using the OTF mapping technique (with the same
reference position), and the
pointing accuracy 
was better than 3\arcsec . 

To produce the final NRO45 map we combined the BEARS and FOREST observations. As a convolution function, we adopted a spheroidal 
function with a spatial grid size of 7\arcsec.5, 
resulting in the final effective angular 
resolution of $\sim$ 22\arcsec. 

As a check on the adopted intensity scale for 
the NRO45 data, comparisons were made with 
published maps of the Orion A cloud from 
the FCRAO 14m telescope \citep{Ripple2013} 
and from the Bell Lab 7m telescope
\citep{1987ApJ...312L..45B}.
We also compared the $^{13}$CO(1-0) NRO45 data with 
recent observations conducted with the IRAM 30m 
telescope (projects 030-15 and 127-15, 
Sanchez-Monge, private communication).
To ensure a meaningful comparison, the NRO45 image 
cube was convolved with a  Gaussian kernel using 
the MIRIAD {\tt convol} command so that there was a 
good match between the spatial and velocity 
resolutions of the other data sets.  
We find that the NRO45 intensity scale is 
about 15 to 25\% higher than those in the FCRAO and Bell Lab 
maps, and within 15\% with respect to the IRAM 30m data.
However, this level of discrepancy is not unusual 
for single-dish observations due to calibration uncertainties. 

\section{Combining NRO45 and CARMA Data}

Here, we provide the practical details of how we
combined the single-dish NRO45 and CARMA
interferometer observations in the uv plane, 
following the procedure outlined in 
\citet{2011ApJS..193...19K}.
The data included in the combination ranges 
from zero-spacing fluxes to measurements on 
the maximum baselines provided by CARMA. 
This is especially important for regions with 
such high spatial dynamic range as the Orion A cloud.

\subsection{Relative Flux Scale}\label{subsubsec:scalefac}

The first step in combining the 
NRO45 and CARMA data is to establish the 
relative flux scale factor,
$f$ $\equiv$ $F_{\rm CARMA}/F_{\rm NRO45}$, 
between the interferometer and single-dish maps.  
The NRO45 telescope 
is treated as an interferometer with baselines 
ranging from 0 to 45 m,
corresponding to uv distances of 0 to 17 k$\lambda$. 
For reference, the complete CARMA 
dataset encompasses baselines from 3 to 40 k$\lambda$. 
The factor $f$ is 
determined by selecting visibilities in the uv space where the
two datasets overlap and comparing the respective fluxes.
In cases where there are deviations 
from the canonical value of $f$=1, 
the interferometer 
flux is adopted because it is typically based 
on more frequent and more reliable calibration observations 
\citep{2011ApJS..193...19K,2013ApJ...774...22P}. 

In practical terms, we are modeling the NRO45 image so that 
it appears to have been observed by a subset of the 
CARMA baselines, i.e., the range of baselines 
included in both maps. This requires some editing of 
each data set so that only regions with significant 
emission in both single-dish and interferometer images 
are included. Since the NRO45 sensitivity on baselines 
longer than 6 k$\lambda$ is relatively low 
\citep[see Figure 5 in][]{2011ApJS..193...19K}, 
the CARMA dataset was limited to observations from 
the 6m-antenna pairs with baselines between 3 and 
6 k$\lambda$. The velocity channel comparisons were 
similarly constrained. CARMA dirty images of these 
``sub-regions'' were produced by Fourier transforming 
the resulting visibilities using MIRIAD. 

For the NRO45 dataset, 
the intensity scale  was converted to 
Jy/beam using Equation \ref{eq:TtoS}
(where $\Omega_{\rm mb}$ = 1.133*FWHM$^2$ with FWHM in radians)
\begin{equation}\label{eq:TtoS}
S_\nu = T_{\rm mb} \frac{2k\Omega_{\rm mb}}{\lambda^2}.
\end{equation}
The resulting image was then regridded to 
match the CARMA map in terms of  
pixel size and channel width, as well as the total 
number of pixels and channels. 
It was also deconvolved using the original 
beam ($\sim 22$\arcsec). This produced a fair 
representation of the NRO45 sky brightness distribution 
to be modeled as if had been observed 
with the same uv sampling
as the modified CARMA map. 
To achieve this, the edited NRO45 image was 
broken down (de-mosaic)
to correspond to the different pointings 
that produced the CARMA mosaic map of the same 
area, using the MIRIAD task {\tt demos}.

As part of this ``de-mosaic'' process the 
response of a primary beam was included at 
each of the pointings to compensate for 
the fact that the MIRIAD task {\tt invert} 
that is used later to produce a dirty image from the 
visibility data automatically corrects 
for the primary beam response 
\citep[see][for details]{2011ApJS..193...19K}. 
Each pointing from the de-mosaic procedure for NRO45 was  
modeled based on the visibility distribution of the 
modified CARMA map described above (using the MIRIAD 
task {\tt uvmodel}). The visibilities generated for each 
pointing were then merged, and the combined set was 
Fourier transformed to produce 
a dirty image of the modified NRO45 dataset (using {\tt invert}). 
As a result, each dirty image is constructed 
from an identical set of baselines, 
selected from the CARMA visibility 
data between 3 and 6 k$\lambda$, 
and has the same dirty beam size.

Ratios of intensities were determined for 
each pixel in the modified CARMA and NRO45 
dirty maps with a signal-to-noise ratio greater than 20. 
Plots of the distribution of the ratio 
for different regions and 
channels show a consistent Gaussian-like 
distribution with an average peak value 
of about 1.6 to 1.8 
for $^{12}$CO, and 1.5 to 1.6 for $^{13}$CO 
and C$^{18}$O respectively. 
We note that the C$^{18}$O(1-0) ratio was 
derived from integrated intensity maps
since the emission is relatively weak, 
especially when the baseline range is 
constrained to 3-6 k$\lambda$.
Adopted flux scale factors are 1.6 for $^{12}$CO(1-0)
and 1.5 for $^{13}$CO(1-0) and C$^{18}$O(1-0), respectively.
These values are unexpected in that they are much 
higher than  the canonical value of $f$ = 1. 
Consequently, we have carefully re-examined 
the flux calibration results and all the 
assumptions and procedures used in deriving this factor.

A comparison of our CARMA intensities with 
other millimeter-wave interferometer measurements 
can help assure the accuracy of our data calibration. 
For example, we considered the $^{12}$CO(1-0) and 
$^{13}$CO(1-0) CARMA data obtained by \citet{2016ApJ...831...36T} 
of a region in Orion in the southern part of our map. 
Both datasets were edited to ensure as much similarity 
as possible between the baselines involved, 
and bright emission structures were compared.
We found the measured intensities 
to be consistent within the noise.
A second comparison used unpublished 
$^{12}$CO(1-0) observations from the ALMA 7m dishes, the 
Atacama Compact Array (ACA) (ALMA project 2016.1.01123.S,
A. Hacar, private communication). We modeled the 
CARMA measurements to appear as if they it were 
observed with the baseline configuration of the ACA. 
Spectra and intensities in different regions of the resulting 
CARMA cube are also consistent with the ACA data. 
Finally, we used the CASA (Common Astronomy Software Applications) 
data reduction package \citep{2007ASPC..376..127M}
to simulate ACA observations of the 
NRO45 image and derived a similarly high flux scale factor 
between NRO45 and the ALMA data. A similar value of $f$ also 
resulted when conducting similar 
tests in MIRIAD with the ACA data.  

Tests of the various procedures involved in our 
derivation of $f$ were also carried out. 
In particular, we repeated the procedures 
described using multiple regions and channels, 
and constraining baselines to different uv ranges 
(e.g., 3-4 k$\lambda$, 4-5 k$\lambda$, 5-6 k$\lambda$, 
3-5 k$\lambda$). We also introduced different 
combinations of the 6m and 10m dishes 
(i.e. different CARMA antenna pairs). 
In all cases, similar values of the 
scale factor were obtained.
To ensure that the flux factor derivation 
is self-consistent, we also
compared the CARMA image with ``itself''. 
Specifically, we smoothed the 
CARMA cleaned image to match the NRO45 resolution, 
and then used the same script to compare the smoothed CARMA
image with the original image. The resulting flux ``ratio''
was 1, as expected. We were able to 
reproduce the original NRO45 data
to within 3\% by imaging the modeled NRO45 visibilities, 
confirming the validity of the 
deconvolution procedures. Using the {\tt immerge} command in
MIRIAD, we again determined the flux scale 
factor and found values of 
$\sim 1.6$ for $^{12}$CO in various sub-regions.

\subsection{Combining the Matched Datasets}\label{subsubsec:combine}
With the NRO45 and CARMA fluxes matched, 
and the NRO45 observations treated as short 
spacing measurements in the larger CARMA 
$uv$ plane, combining the two datasets is 
already more straightforward. The CARMA 
dataset requires no further overall 
modification and a CARMA-only dirty map 
can be immediately constructed. For this 
preliminary map, we used the  complete set 
of visibilities from the region mapped by 
CARMA but restricted to the velocity range 
0 to 16.7 km s$^{-1}$. Velocity channel 
widths are 0.25 km s$^{-1}$ for $^{12}$CO(1-0), 
and 0.22 km s$^{-1}$
for both $^{13}$CO(1-0) and C$^{18}$O(1-0). 
For each pointing in the mosaics, 
we adopted a cell size of
2\arcsec\ and an image size of 95 cells,
more than twice the size of the synthesized 
beam of the CARMA 6-m telescopes. 
These choices were largely dictated 
by our imaging requirements, i.e., pixel size 
$\sim$ 1/5-1/3 $\times$ synthesized beam; 
image size $>$ 2 $\times$ field of view.
Smaller cell sizes or larger image sizes 
require considerably more computing resources 
in terms of memory and time. The resulting 
CARMA-only dirty map was produced with the 
MIRIAD {\tt invert} task, adopting a robust 
number of 0.5 and {\tt options=mosaic,double}.  
 
By contrast, a few more steps are necessary 
to prepare the modified and flux-corrected 
NRO45 visibilities for combination because
we will work on the entire NRO45 image cube
in order to fill the uv ``hole'' in the CARMA visibilities.
First, the NRO45 cube was converted from 
main-beam temperature $T_{\rm mb}$ 
to flux density $S_\nu$
using Equation \ref{eq:TtoS}. Then, we multiplied
the new cube by the flux scale factor 
(see section \ref{subsubsec:scalefac}). 
The {\tt maths} task was used for the two steps.
Next, we regridded the NRO45 image cube to match 
the pixel and channel numbers and sizes of the 
CARMA-only preliminary dirty image, and deconvolved it 
with the NRO45 beam ($\sim 22$\arcsec), 
using the MIRIAD command {\tt convol} with 
{\tt options=divide}. The de-mosaic procedure described in 
section \ref{subsubsec:scalefac} was then applied.
In this way, we obtain visibilities at each pointing 
of the CARMA mosaic as if the observations had been made 
with a set of randomly generated baselines that 
follow a Gaussian distribution and mimic 
the response of the NRO45 telescope. 

The distribution of baselines was generated using, 
{\tt hkuvrandom}, a modified version of the standard task 
{\tt uvrandom} developed by J. Koda in MIRIAD 4.3.8. 
In effect, a Gaussian distribution of uv points (baselines) 
with a standard deviation {\tt sdev} is generated.
This Gaussian is the Fourier transform of the NRO45 
primary beam with a standard deviation of 3.5 k$\lambda$. 
Due to our slightly larger effective primary beam, 
22\arcsec, this is slightly smaller than the value 
(3.9 k$\lambda$) used in \citet{2011ApJS..193...19K}.

The total number of points generated by {\tt hkuvrandom}
is given by the parameter $N_{\rm total}$ 
({\tt npoint} in {\tt hkuvrandom}). 
It depends largely on the value assumed 
for the NRO45 visibility integration ($t_{\rm vis}$, 
{\tt inttime} in {\tt hkuvrandom}) and on the requirement 
that the sensitivity per pixel in $uv$ space 
for NRO45 visibilities is similar to that for CARMA 
in their $uv$ overlap range, 
3-6 k$\lambda$ \citep{2011ApJS..193...19K}. 
Following \citet{2011ApJS..193...19K}, we adopt
$t_{\rm vis}$ = 0.01 s. This is sufficiently low to ensure that
$N_{\rm total}$ is large enough to fill the uv space. 
We adopted $N_{\rm total} \sim$ 3000. 
Due to the low sensitivity of NRO45 observations 
on baselines greater than 6 k$\lambda$, we flagged 
such generated baselines before we applied the 
{\tt uvmodel} MIRIAD command to the de-mosaicked 
images and used the remaining generated baselines 
to produce a visibility file for each mosaic pointing.
The file header for each file was modified to 
include the NRO45 system temperature, the weighting parameter
{\tt jyperk} \citep[see][equations (14)(15)]{2011ApJS..193...19K}, 
and our arbitrary 2\arcmin\ Gaussian primary beam
(chosen to be larger than CARMA primary beams). 
Finally, we merged all the modeled NRO45 visibility files together
to combine with the CARMA visibilities.

Imaging of the combined data was carried out with the Yale 
high performance computing (HPC) clusters, Grace and 
Farnam\footnote{\url{https://research.computing.yale.edu/services/high-performance-computing}}.
We produced a dirty cube of the joint CARMA and NRO45 
observations by Fourier transforming the combined 
visibilities using the task {\tt invert}. 
This requires roughly 40 GB of memory. 
As before, we adopted a 2\arcsec\ pixel size, 
approximately 1/5-1/3
of the final beam size, and an image size at 
each mosaic pointing of $\sim$ 95 pixels. 
The invert process for the 
$^{12}$CO cube ($\sim$ 2500$\times$4300 pixels, 90 channels)
took roughly 1 day and 2.5 GHz CPU.
Each channel was cleaned separately, 
using the {\tt mossdi2} task in MIRIAD.
For this procedure, each channel took up 1 CPU 
and roughly 30 GB memory (less memory is needed for 
channels with lower emission). All channels can be 
cleaned simultaneously using multiple CPUs. 
The cleaning process, with the stopping threshold 
for clean set to 3 times the RMS noise level,
took at most one day for each channel. 
The beam size and Position Angle, 
channel width,
and RMS noise per channel of the 
resulting maps are given
in Table \ref{tab:sensitivity}.

\begin{figure*}[htbp]
\epsscale{1.}
\plotone{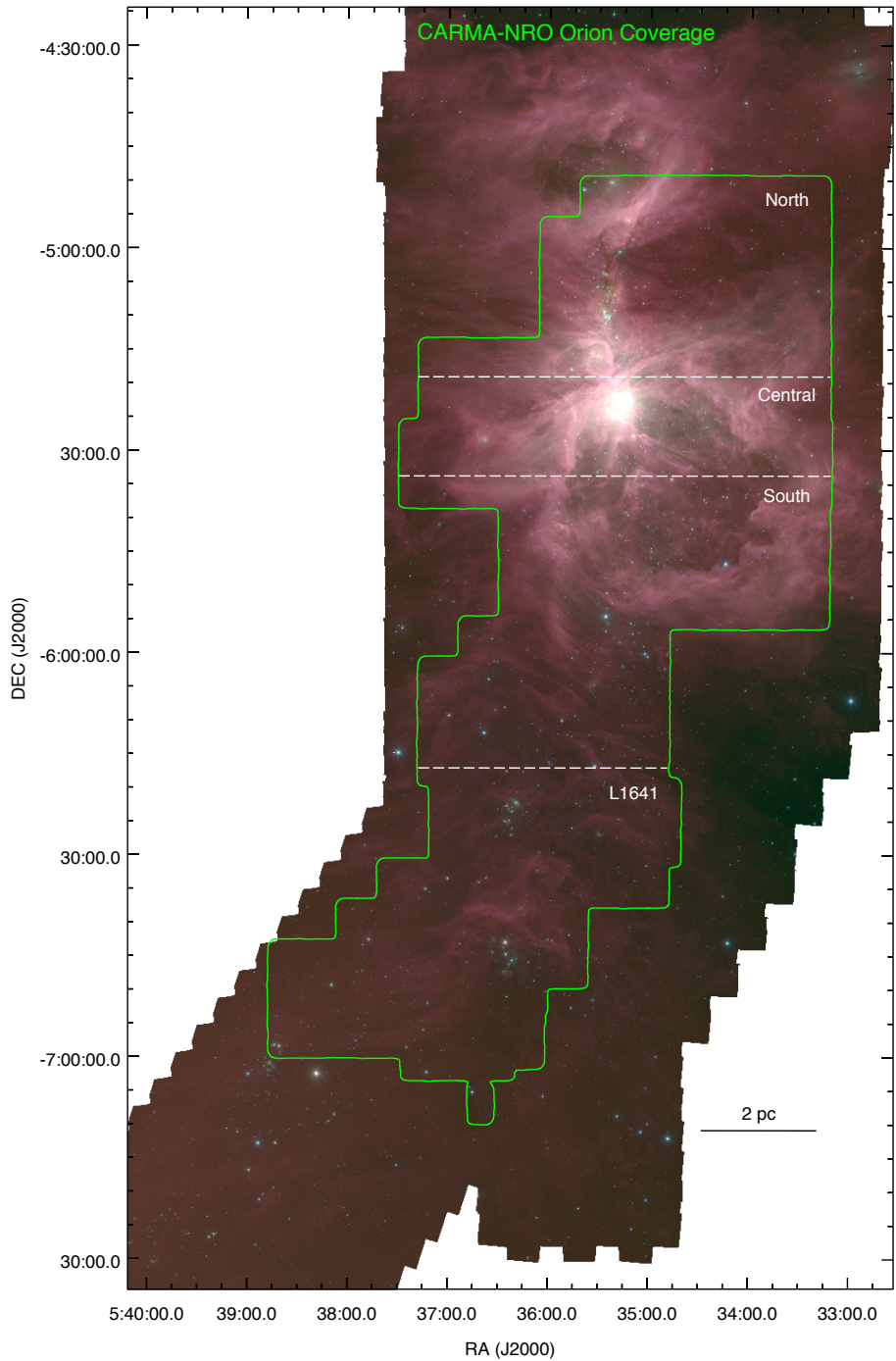}
\caption{
The footprint of the $^{12}$CO CARMA-NRO Orion map 
(green solid lines) overlaid on an Spitzer/IRAC false
color RGB image based on 8$\mu$m (Red), 4.5$\mu$m (Green), 
and 3.4$\mu$m (Blue) observations. White dashed lines 
designate four regions of interest (Feddersen et al., 
submitted) that will be used in Section \ref{subsec:kin}.
\label{fig:coverage}}
\end{figure*}

Figure \ref{fig:coverage} shows the entire mosaic footprint for 
the $^{12}$CO combined image overlaid on a RGB false color 
image from Spitzer (Rob Gutermuth, private communication). 
The footprints of the $^{13}$CO and C$^{18}$O 
combined images are progressively smaller
(due to limited coverage in NRO45 observations).
Most importantly, all the maps span slightly more 
than 2 degrees, or about 14 pc,  in declination
and encompass a variety of well-known star-forming
regions in the Orion cloud.

\begin{figure*}[htbp]
\epsscale{0.41}
\centering
\hspace{-18pt}
\plotone{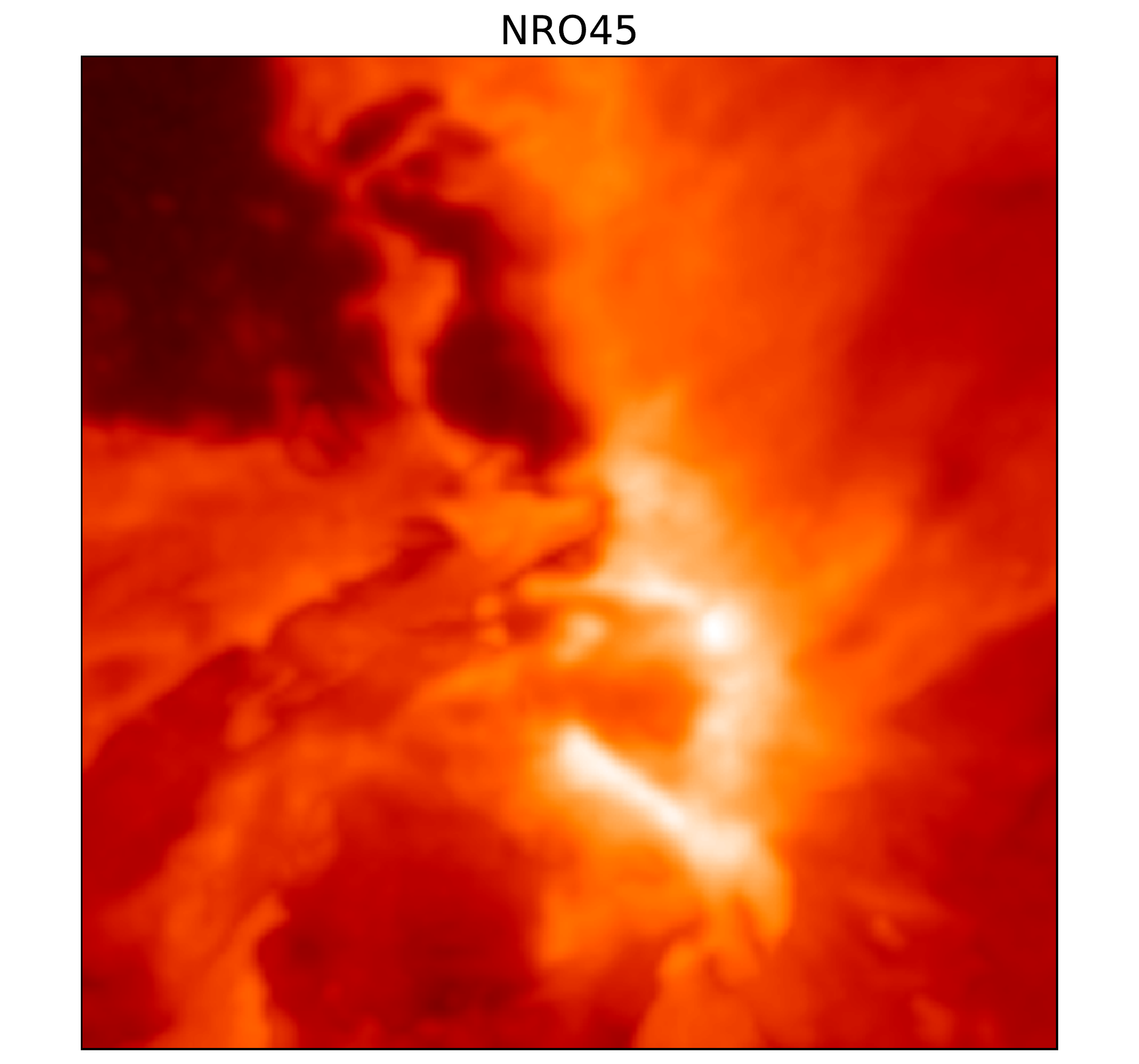}
\hspace{-18pt}
\plotone{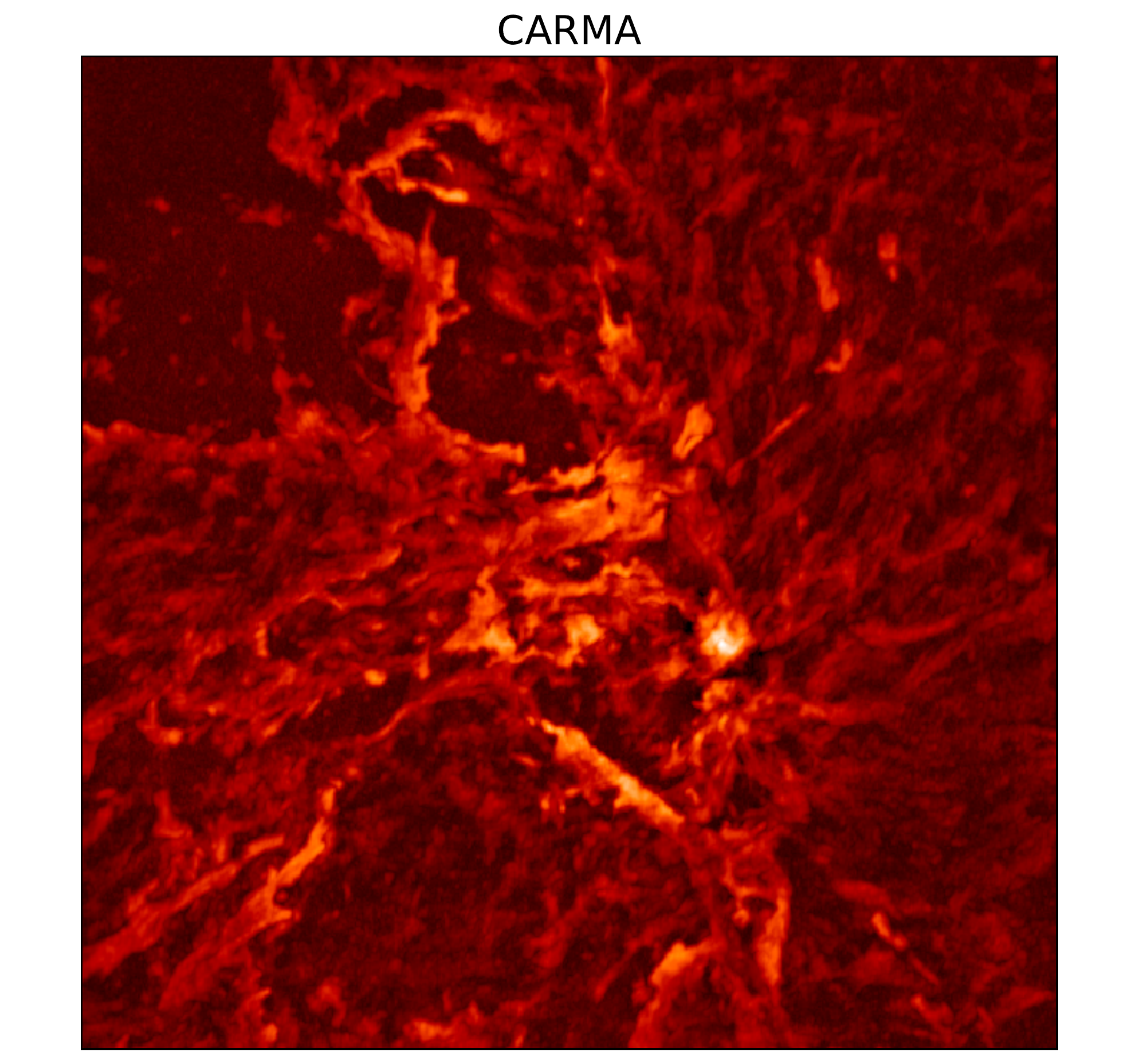}
\hspace{-18pt}
\plotone{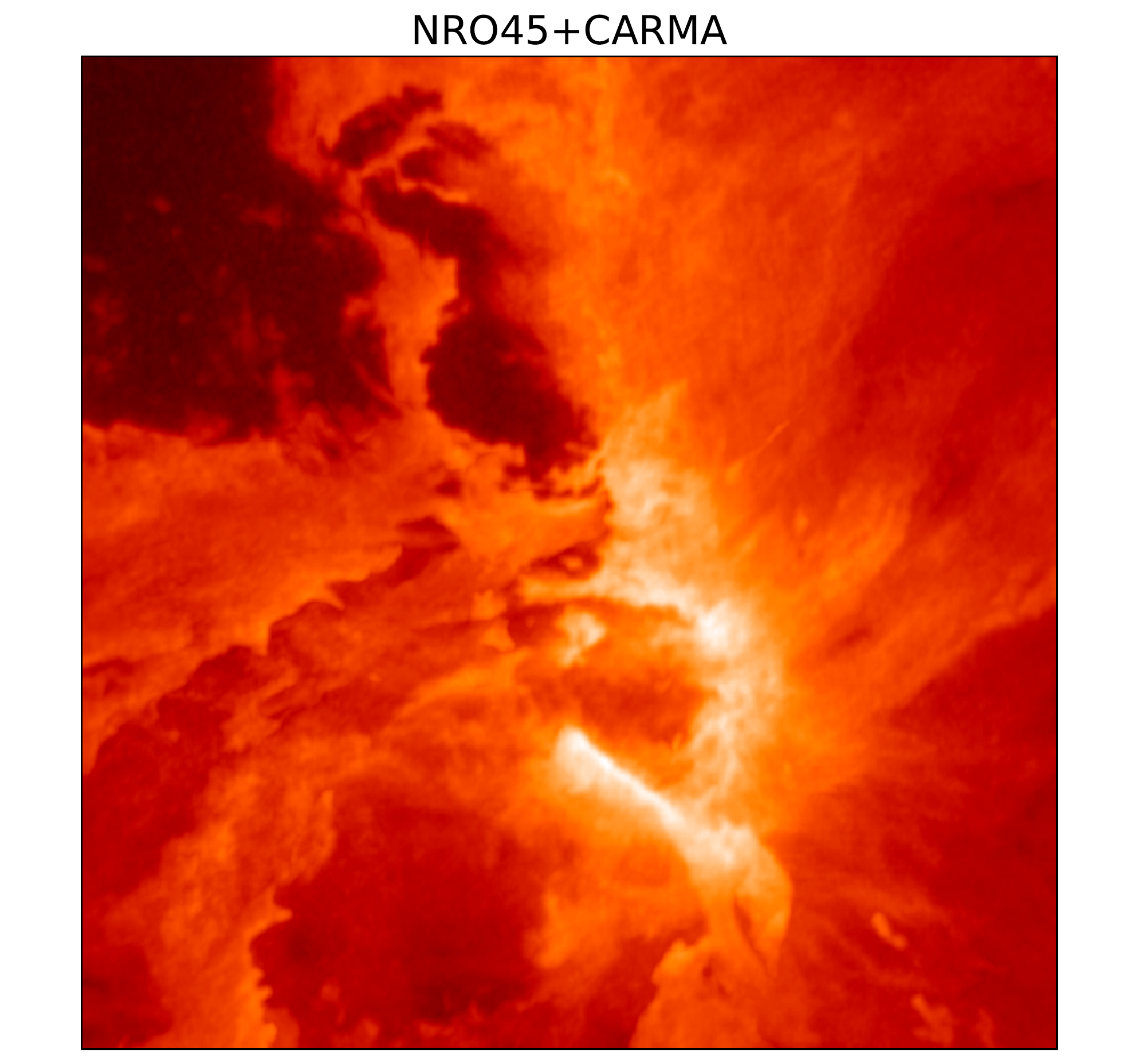}
\hspace{-18pt}
\caption{Sample comparison between NRO45, CARMA and combined maps.
 ({\it Left:}) $^{12}$CO peak intensity maps from NRO45, ({\it Middle:}) CARMA, ({\it Right:}) and the combined CARMA+NRO45 datasets. All panels show the same  19.0\arcmin\ by 18.7\arcmin\ 
(2.21 pc by 2.17 pc) area centered around the ONC and Orion KL region and includes the Orion Bar. 
The increased sharpness of the extended images when the CARMA observations are combined with those from NRO45 is immediately clear. 
\label{fig:combine}}
\end{figure*}
Figure \ref{fig:combine}, a comparison of the $^{12}$CO peak 
intensity maps of a sub-region in the Central Orion
box (see Figure \ref{fig:coverage}) based on
NRO45, CARMA, and the combined NRO45+CARMA datasets
demonstrates the effectiveness of combining single dish and 
interferometer observations. The 19\arcmin\ x 18.7\arcmin\
(2.21 x 2.17 pc) maps are centered on the dense part of the 
Orion Nebula Cluster (a.k.a., the Huygens Region, which hereafter we denote as ONC) and the 
Orion KL regions at R.A. $\rm 05^h35^m26\fs6$, 
DEC. $\rm -05\arcdeg20\arcmin44.9\arcsec$. 
There is a dramatic improvement in the image quality
after the CARMA and NRO45 data are combined, 
with considerably more 
detailed structure visible over the extended sub-region. 
It is worth noting that, over this sub-region, 
CARMA recovers only about 20\% of the NRO45 detected flux.

\begin{deluxetable}{ccccc}
\tablecolumns{5}
\tablewidth{0pt}
\tablecaption{Final sensitivity \label{tab:sensitivity}}
\tablehead{
\colhead{Transition} &
\colhead{Beam} &
\colhead{PA} &
\colhead{$\Delta_{V}$} &
\colhead{$\sigma_{K}$}\\
\colhead{} & 
\colhead{} & 
\colhead{(deg)} & 
\colhead{$(\rm km~s^{-1})$} & 
\colhead{(K)}}
\startdata
$^{12}$CO(1-0) & $10\arcsec\times8\arcsec$ & -13 & 0.25 & 0.86 \\  
$^{13}$CO(1-0) & $8\arcsec\times6\arcsec$ & 10 & 0.22 & 0.64 \\    
C$^{18}$O(1-0) & $10\arcsec\times8\arcsec$ & -0.4 & 0.22 & 0.47 \\
\enddata
\tablecomments{The RMS per channel shown in the last column is the median value across the relevant cube.}
\end{deluxetable}

\subsection{Using $\Delta$-variance analysis to verify data}

A $\Delta$-variance analysis \citep{Stutzki1998,Ossenkopf2008} 
can also provide confidence in the quality
of the final emission line 
images. Such analysis readily reveals 
peculiarities and artifacts in the spatial distributions 
and, at the same time, can characterize
the global scaling behavior and prominent 
spatial scales in the maps. In principle, 
the $\Delta$-variance spectrum measures the 
relative amount of structure on
a given scale $s$ by convolving the 
map being considered with a radially symmetric 
top-down wavelet of that scale and 
computing the variance of the 
convolved map as a function of $s$. Each data
point can be weighted by a significance 
value before selecting the statistics
to be examined, e.g., variable noise in the data. 
\citet{Stutzki1998} have
shown that for rectangular maps with 
equal weights of all points the 
$\Delta$-variance spectrum is equivalent 
to the power spectrum; the slope, $d$, 
of the $\Delta$-variance spectrum 
corresponds to a power spectral slope
$\beta=d-2$. 

$\Delta$-variance 
analysis can also be applied to maps 
with irregular boundaries and
data with variable uncertainties \citep{Ossenkopf2008} 
to obtain statistics on the 
size dependence of all structures in
the map. Here, we applied the analysis to the integrated
intensity maps for each spectral line 
and also to maps of the intensity in
the central velocity channel at
$v=8.8$~km s$^{-1}$.
For our statistical 
analysis we weighted each
data point by the inverse RMS noise in the 16 
channels at the velocity boundaries of the cubes.

\begin{figure}
\centering
\includegraphics[width=0.95\columnwidth]{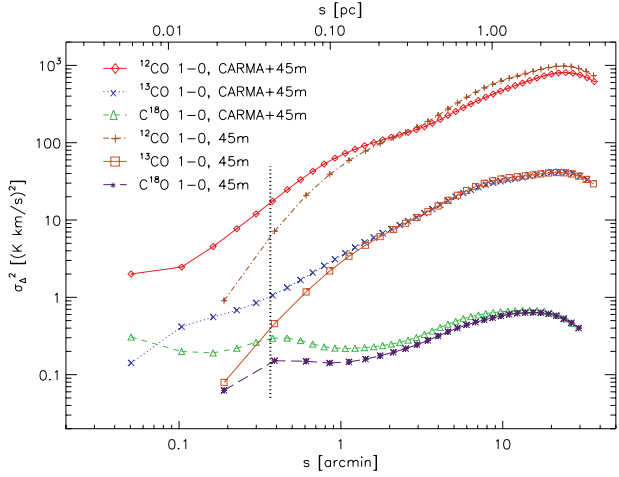}
\caption{$\Delta$-variance spectra for the
$^{12}$CO(1-0), $^{13}$CO(1-0), and C$^{18}$O(1-0)
integrated intensity maps from NRO45 and
from the combined (CARMA+NRO45) dataset.
On the x-axis, the lag indicates the scale
measured by the wavelet size. The y-axis 
displays the variance of the structural 
fluctuations at each value of x in units 
of the square of the measured map. 
The vertical dotted line indicates the
NRO45 beam size}.
\label{fig:Dvar1}
\end{figure}

\begin{figure}
\centering
\includegraphics[width=0.95\columnwidth]{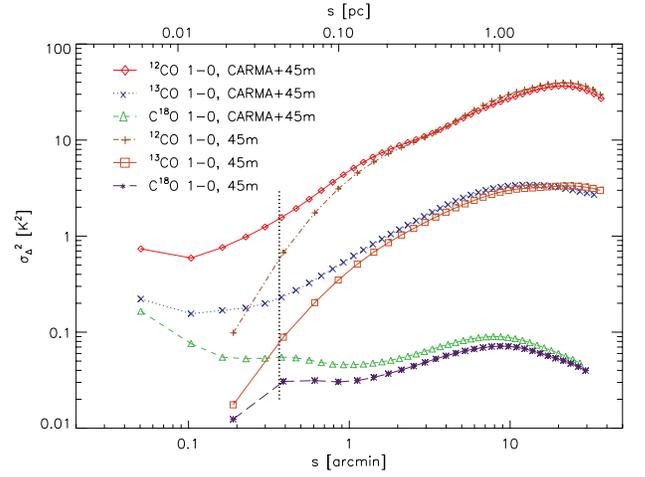}
\caption{ $\Delta$-variance spectra for maps of
the central velocity channel intensity 
($v=8.8$~km s$^{-1}$ with channel width
$\Delta v=0.25$ for $^{12}$CO, 
and 0.22 km s$^{-1}$  for $^{13}$CO and C$^{18}$O.}
\label{fig:Dvar2}
\end{figure}

Figure~\ref{fig:Dvar1} shows the 
$\Delta$-variance spectra for the integrated intensity maps.
Figure~\ref{fig:Dvar2} shows the corresponding
spectra for the channel maps. 
Both demonstrate that  combining
NRO45 and CARMA data introduced no 
obvious artifacts in the $^{12}$CO and  
$^{13}$CO maps. Both cubes show a perfect power law at
scales around 22\arcsec\, the effective NRO45 beam size.
For C$^{18}$O, the agreement was not as good.
In addition, the $\Delta$-variance spectrum 
displays a peak at about 25\arcsec\ and an 
inspection of the C$^{18}$O map indicates 
the presence of granular structure on that scale. 
Both results are likely attributable to overall 
lower signal-to-noise in the C$^{18}$O maps,
greater noise in the NRO45 maps than in CARMA maps, 
and accompanying issues with the cleaning process. 

At lag scales below 0.13\arcmin\ (8\arcsec)
the $\Delta$-variance spectra of all maps
are dominated by the 
natural limitations of the CARMA
observations. Since the major axis of the 
synthesized CARMA beam is
about 8\arcsec, structures with smaller 
sizes will be blurred, reducing their contribution to the 
$\Delta$-variance spectrum. 
In contrast, $\Delta$-variance spectra tend to increase 
at the smallest lags since the small-scale spectrum is 
dominated by observational noise that introduces excess power. 
This effect is particularly relevant to 
the central velocity channel
maps where the relative noise is higher 
than in the integrated intensity maps.
The impact of the finite telescope beam
can be seen when the spectra for the full combined  
data set is compared with that obtained
from the NRO45 maps at the 22\arcsec\ 
resolution. Structural variations on scales up
to about three beam widths are suppressed 
in the absence of the CARMA observations as expected from 
computations by \citet{2001A&A...366..636B}.
At larger scales the $\Delta$-variance 
spectra show no resolution
dependence. Likewise, smoothing the full data 
set to the resolution of the
45m telescope data reproduced the spectra
obtained from the single-dish maps.

\section{Results}\label{sec:results}

\subsection{Gas Distribution}\label{subsec:gasdistr}

\begin{figure*}[htbp]
\epsscale{1.1}
\plotone{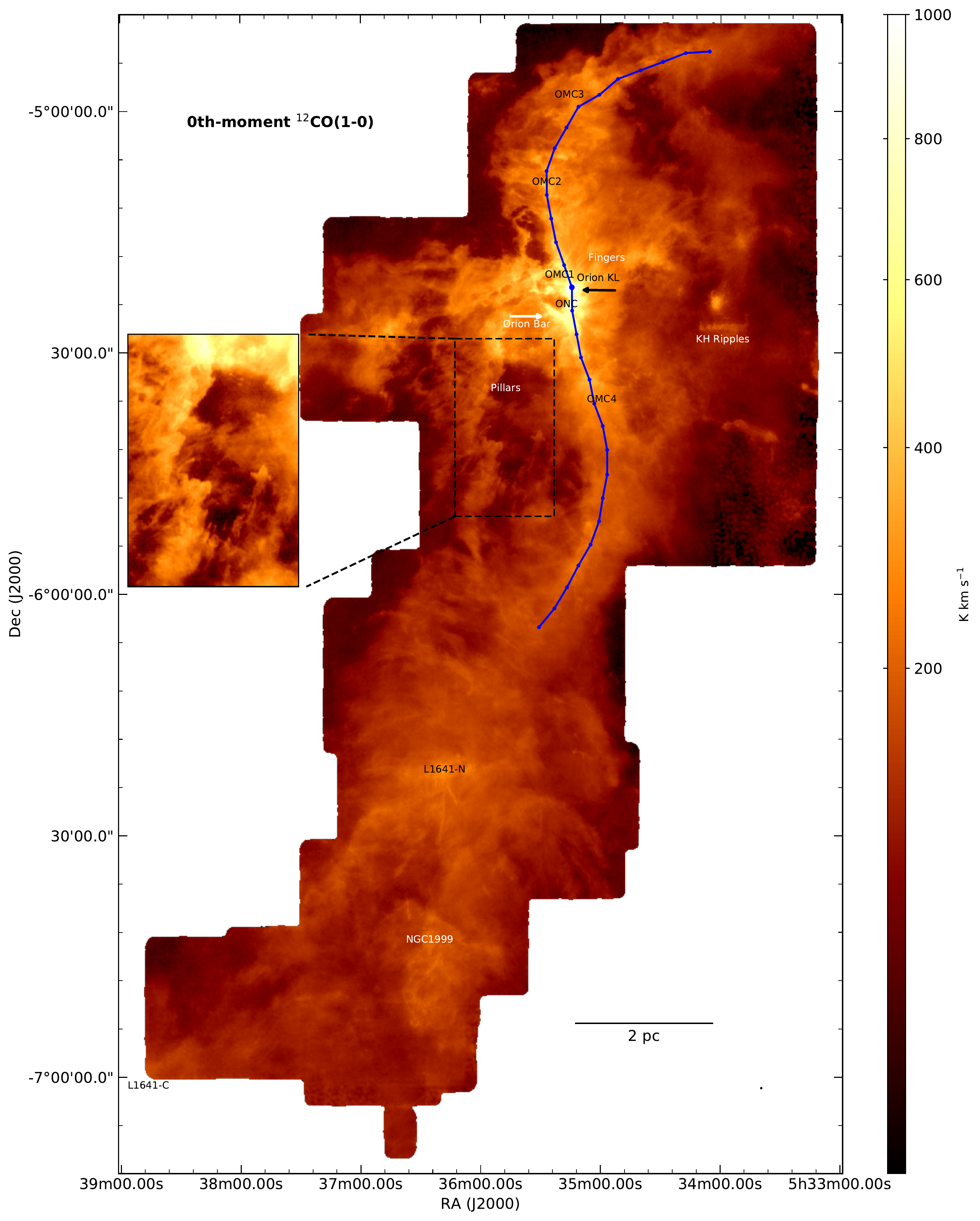}
\caption{
0th-moment (integrated intensity)
map of the $^{12}$CO(1-0) emission over the
velocity range 2.5 to 15 km s$^{-1}$. 
No intensity clipping has been applied. 
The color bar on the right provides the square root
intensity scale in units K $\rm km~s^{-1}$. 
A zoom-in view of the Pillars is displayed in the inset.
Noisier patches seen near R.A. $\rm 05^h33^m25\fs0$ 
at various declinations result from 
limited NRO45 observing time.
The blue curve is the cut for the PV diagram
defined in Section \ref{subsec:kin}.
The blue dots indicate 3\arcmin\ intervals along the curve.
\label{fig:12mom0}}
\end{figure*}

\begin{figure*}[htbp]
\epsscale{1.1}
\plotone{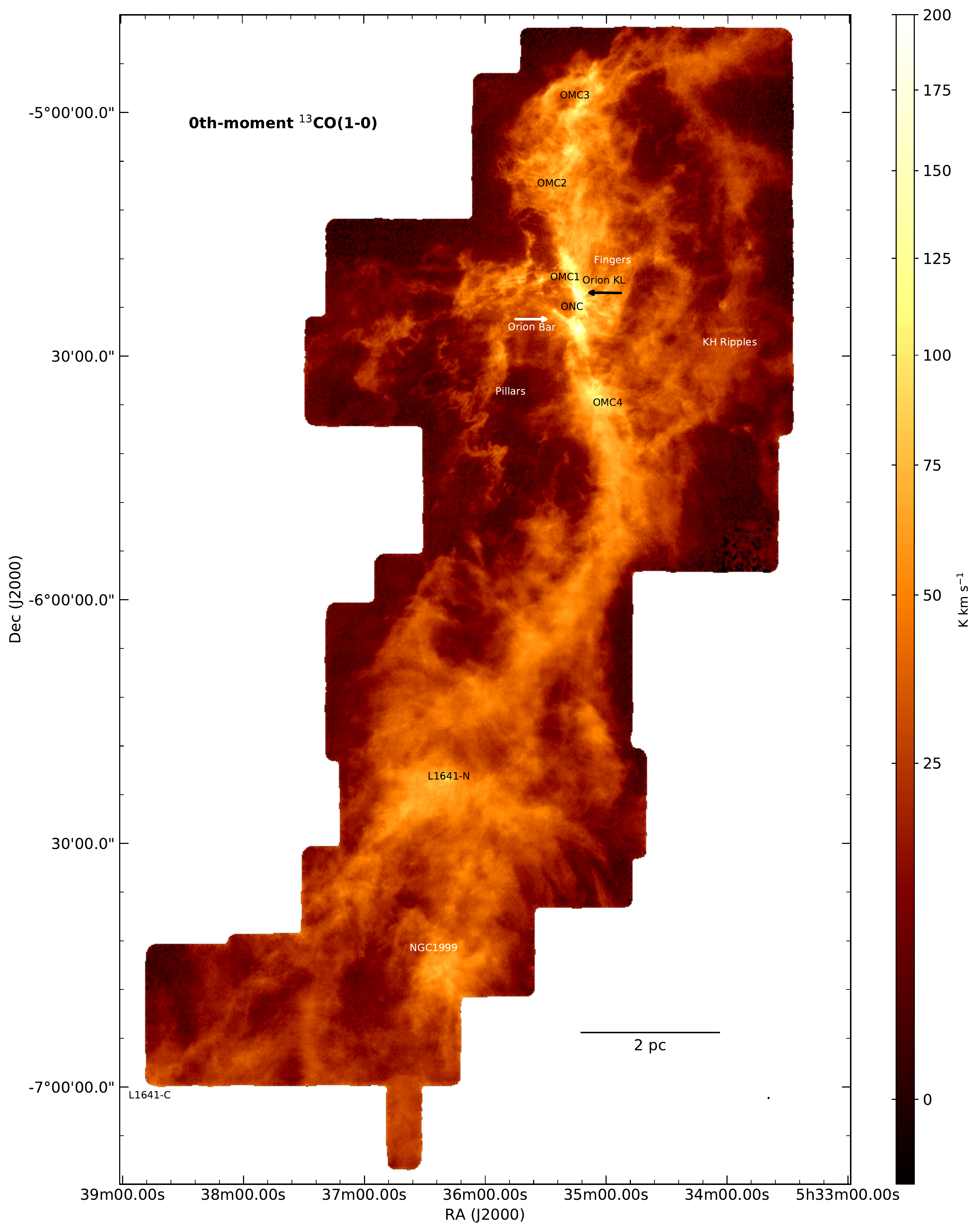}
\caption{
As Figure \ref{fig:12mom0} but for $^{13}$CO(1-0) emission.
\label{fig:13mom0}}
\end{figure*}

\begin{figure*}[htbp]
\epsscale{1.1}
\plotone{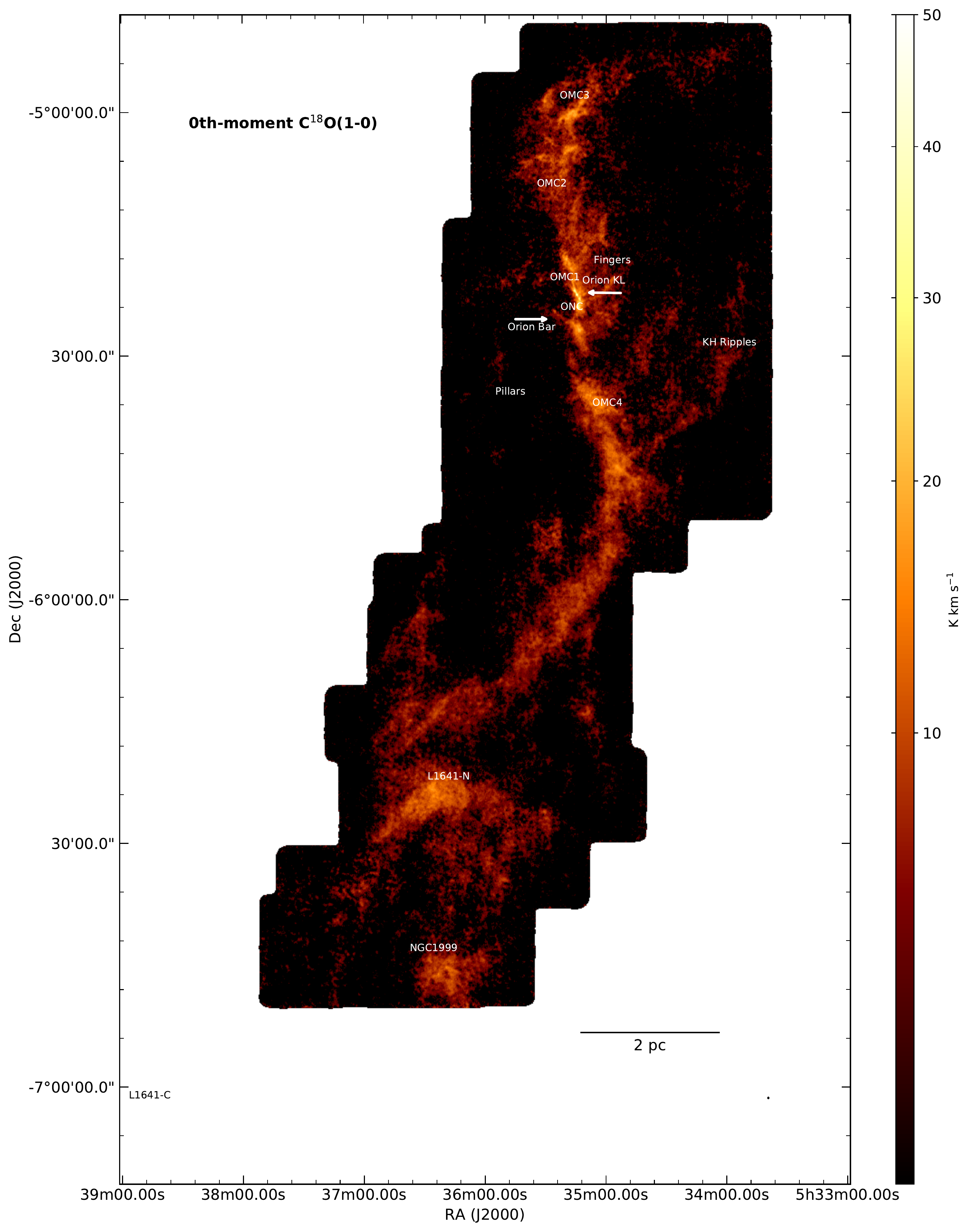}
\caption{
As Figure \ref{fig:12mom0} but for C$^{18}$O(1-0) emission.
\label{fig:18mom0}}
\end{figure*}

Figures \ref{fig:12mom0}, \ref{fig:13mom0}, and \ref{fig:18mom0}
show the integrated intensity maps\footnote{These and most other maps in this paper
are available online at: https://dataverse.harvard.edu/dataverse/CARMA-NRO-Orion} 
for $^{12}$CO(1-0), $^{13}$CO(1-0),
and C$^{18}$O(1-0), 
respectively, over the velocity range 
2.5 to 15 km s$^{-1}$.
As already noted, the maps span about 2 degrees 
in declination and encompass a range of star-forming
regions. Each extends from  OMC-3 
and OMC-2, at declination $\sim$ -5\arcdeg,  to   
the V380 Ori Group (NGC 1999) and the L1641-C 
region at nearly  $\sim$ -7\arcdeg.
Several well-known regions are labeled.
The relatively noisy patches that are noticeable 
around R.A. $\rm 05^h33^m25\fs0$, particularly in
the $^{12}$CO(1-0) map, are due to limited NRO45 
observing time and resulting lower sensitivity. 
Other widespread, complex structures can also be seen.
Some of these have been studied extensively but many
have barely been mentioned in the literature. 
Since we expect numerous detailed studies and analyses 
will follow our dissemination 
of this data, we draw attention to only a few highlights of the new maps here.

The ONC  is probably the most studied region 
in the Orion A cloud \citep[e.g., Figure 1 of][]{Meingast2016}. 
Not surprisingly, it appears as one of the most dramatic 
regions in all three of our molecular line maps.  
In particular, the area enclosing the ONC is clearly 
seen as a cavity in the gas distribution, 
presumably due to clearing of molecular 
gas in the HII region that is 
powered by high-mass stars in the associated cluster 
\citep{2017ApJ...837..151O}. The cavity itself is bounded
by the bright edges that are expected when 
ionization/dissociation fronts interact with the 
surrounding molecular cloud. The brightest edge, 
to the south-east, is commonly referred to as the 
Orion Bar \citep[see, e.g.,][]{2009ApJ...693..285P,2017arXiv170708869G}. 
Another cavity in the molecular gas distribution 
is visible north of the ONC. 
This is produced by the HII region, M43
\citep{1978A&A....65..207T,2011PASJ...63..105S,2013ApJ...774L..20S}.

Further north, our maps show the molecular gas associated 
with the OMC-2 and OMC-3 star-forming regions. 
These host a wealth of phenomena associated with 
early star formation and include protostars and 
outflows in a range of evolutionary states. 
\citep{2008ApJ...683..255S,2009PASJ...61.1055S,Megeath2012,2013ApJ...767...36S,2013A&A...556A..62L}. 
Immediately to their west, a region of significantly
lower intensity is particularly noticeable in 
the $^{13}$CO map (Figure \ref{fig:13mom0}). 
This is bounded on its far side by a prominent 
arc of emission that continues southwards to OMC-4, 
suggesting the presence of an extended cavity-like structure. 
A similarly elongated, lower intensity region 
is seen to the southeast of the ONC and Orion Bar. 
This depression is bounded on the east by  
structures that have the appearance of irradiated
and elongated globules or pillars 
with bright edges. They can be discerned more clearly in the zoom-in inset 
in Figure \ref{fig:12mom0}, and appear to be pointing 
towards the ionizing stars of the ONC, 
consistent with photo-erosion by the UV-radiation 
emanating from the massive stars in the region. 

One of the most striking structures in the Orion A 
cloud is the so-called integral-shaped filament (ISF) 
first seen in lower resolution maps \citep{1987ApJ...312L..45B}.
In Figure \ref{fig:12mom0} this structure is also detected and stretches about 1\arcdeg\
from north to south (DEC -5\arcdeg\ to DEC $\sim$ -6\arcdeg).
As we discuss below,
considerable filamentary structure, on a variety of scales, 
is especially clear in our $^{13}$CO maps covering different velocity 
ranges and will provide a wealth of information 
for comparison with numerical simulations of 
molecular clouds (e.g., Suri et al., in prep.).

The L1641 cloud lies  
\citep[see][]{2008hsf1.book..621A} at the south end of our maps.
Our maps encompass the gas associated with the L1641-N and  
NGC 1999 clusters, while their southern border 
coincides with the northern outskirts of the L1641-C 
cluster. This region of the Orion A cloud harbors no high-mass 
stars but a few intermediate-mass objects are present, 
as are several groups of young low-mass stars.  
Again, there is considerable potential for a wide 
range of star formation studies (e.g., Arce et al., in prep.).    

\subsubsection{Cloud structure from $\Delta$-variance analysis}

The $\Delta$-variance spectrum of 
the integrated intensity $^{13}$CO map 
(Figure~\ref{fig:Dvar1})
is an almost perfect power law spectrum over
all scales from 20\arcsec\ to 7\arcmin. 
The slope is 1.1, corresponding to a
power spectral index $\beta=3.1$. 
At larger scales the spectrum becomes
shallower, indicating a relative lack of structure. 
For the corresponding channel map (Figure~\ref{fig:Dvar2}), 
the inertial range starts only at 
25\arcsec\ due to effects of higher noise, 
while the power-law
slope is somewhat shallower, $\beta=3.0$
\citep[see discussions in][]{2007ApJ...658..423K,2013ApJ...771..123B}.
Similarly, the integrated intensity
map peaks at about 2.5~pc but the channel 
map shows a peak at a scale of 
1.5~pc (10\arcmin\ corresponds to $\sim 1$ pc 
at a distance of 400 pc).
The C$^{18}$O maps show the same qualitative behavior, 
with somewhat smaller scales, 1.8 and 1.0~pc, implied.
However, the slope cannot be measured precisely
due to higher noise and imaging artifacts. 
We therefore describe the 
structure as self-similar at 
scales up to 0.8~pc and note a relative 
lack of structure at increasingly
larger scales, with only the ISF at the scale of the map.

Applying the same arguments to the channel map spectra of
Figure \ref{fig:Dvar2} enables us to determine the maximum size of 
velocity-coherent structures in our datasets.
While on large scales, the contributions of individual 
filaments can combine to present the appearance of extended,
connected structure, our measurements of optically 
thin lines show that such
connections do not always hold in velocity space. 
The steeper spectrum for our integrated
intensity maps is consistent with a 
positive size-linewidth relation creating 
relatively brighter structures at larger
scales when integrating over 
the full velocity profile.

At scales above 2-3\arcmin, 
the $^{12}$CO data follows the same trend as $^{13}$CO,
showing a short power-law slope with 
the same exponents. However, for $^{12}$CO, the peaks of the
$\Delta$-variance spectra in Figures \ref{fig:Dvar1} and 
\ref{fig:Dvar2} are shifted 
to even larger scales of 2.8 and 2.4~pc, respectively,
as might be expected 
in view of the higher optical depth  
and the sensitivity to somewhat
more extended emission. In addition, 
both maps show a relative excess of
structure at a scale of about 1\arcmin, 
corresponding to 0.1~pc. This is
seen as a prominent bump in the $^{12}$CO
integrated intensity spectrum.
The bump in this optically thick line
is most likely attributable to the
presence of large velocity gradients and
line widths in the contributing regions.
The most prominent broad line region in our maps
is Orion KL, followed by some structures south of
the Orion Bar. All are approximately 1\arcmin\
in size. It is worth noting that if a region of a few
arcmin around Orion KL is masked out, the strength
of the bump in the $\Delta$-variance spectra is 
reduced by 75\%. The combination of all
$\Delta$-variance spectra is thus consistent 
with a large-scale turbulent cascade that is 
locally modified by dynamical feedback
at a scale of about 0.1 pc (1\arcmin).

\subsection{Gas Kinematics}\label{subsec:kin}
\begin{figure*}[htbp]
\epsscale{1.2}
\vspace{-1.5cm}
\plotone{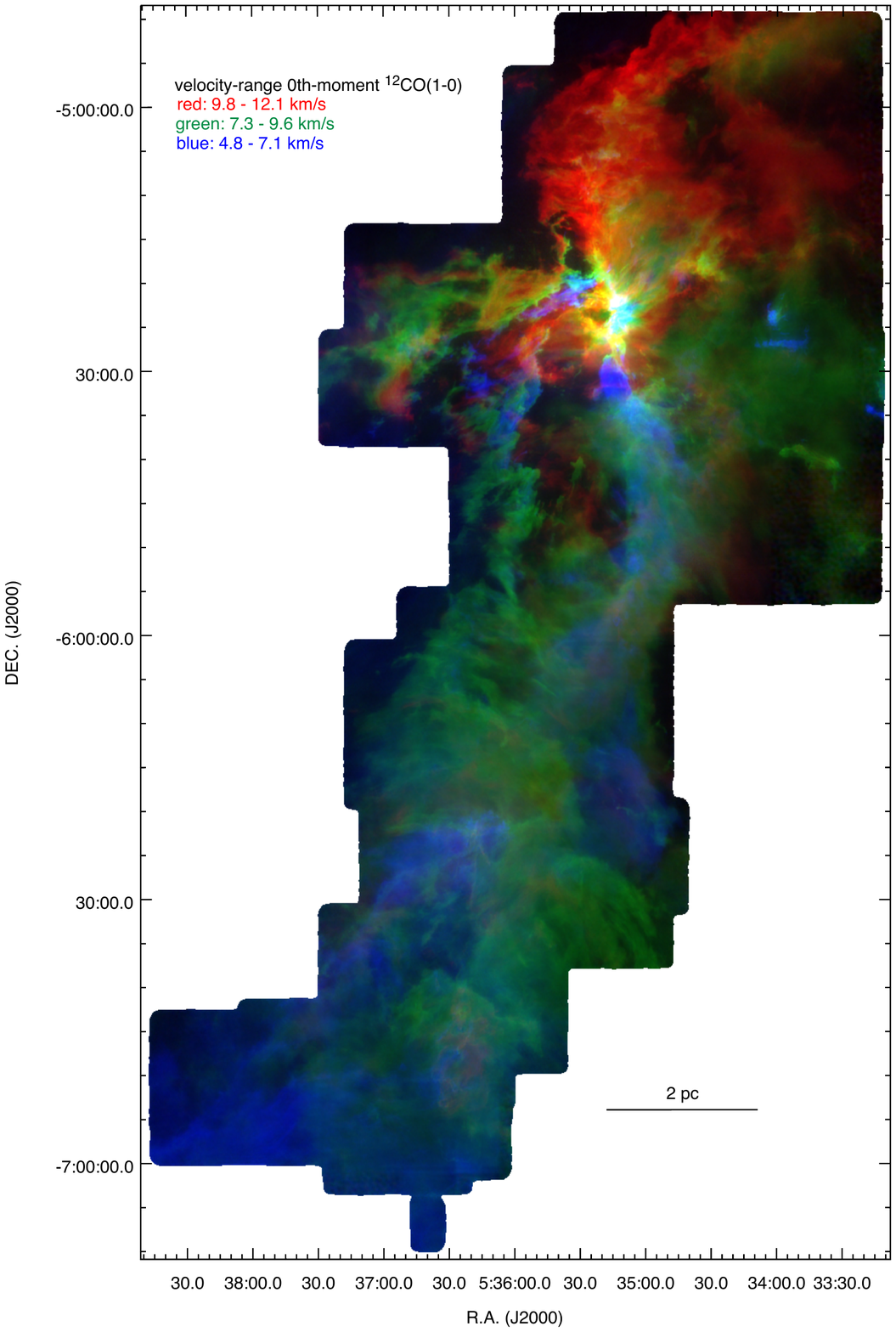}
\vspace{-1.5cm}
\caption{
RGB false color image of velocity-limited, 
integrated intensity maps
for $^{12}$CO(1-0). Integration ranges and colors are:
4.8 - 7.1 $\rm km~s^{-1}$ (blue);
7.3 - 9.6 $\rm km~s^{-1}$ (green);
9.8 - 12.1 $\rm km~s^{-1}$ (red).
\label{fig:12mom0color}}
\end{figure*}

\begin{figure*}[htbp]
\epsscale{1.2}
\vspace{-1.5cm}
\plotone{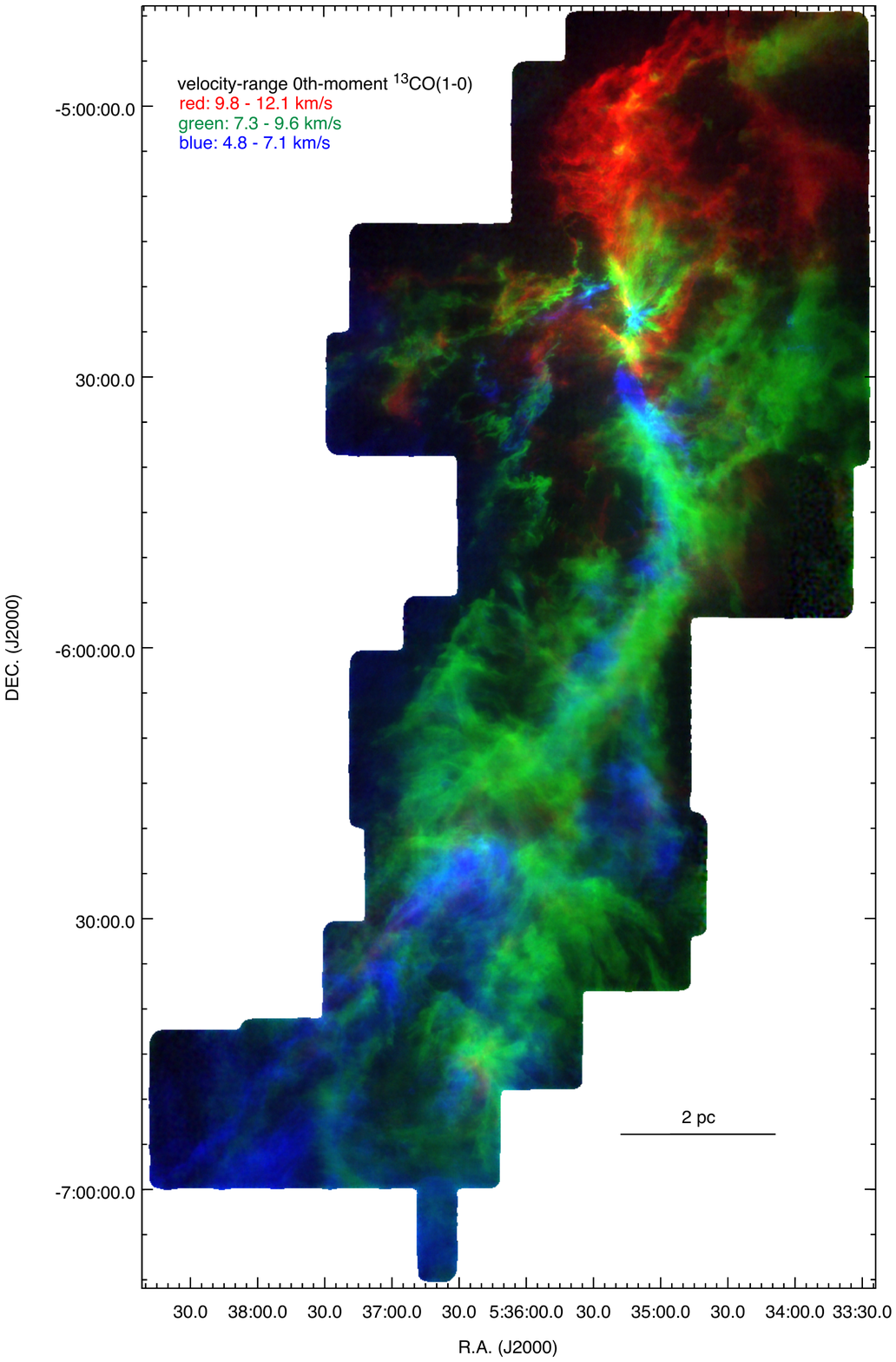}
\vspace{-1.5cm}
\caption{
As Figure \ref{fig:12mom0color} 
but for $^{13}$CO(1-0).
The three integration ranges and colors are:
4.8 - 7.1 $\rm km~s^{-1}$ (blue);
7.3 - 9.6 $\rm km~s^{-1}$ (green);
9.8 - 12.1 $\rm km~s^{-1}$ (red).
\label{fig:13mom0color}}
\end{figure*}

As a first step in understanding the large-scale kinematics
of the Orion-A cloud, we created false color images of 
the integrated intensity maps for $^{12}$CO(1-0) and 
$^{13}$CO(1-0) by limiting the integration velocities
to three broad ranges, 9.8 - 12.1 $\rm km~s^{-1}$ (Red),
7.3 - 9.6 $\rm km~s^{-1}$ (Green), 
and 4.8 - 7.1 $\rm km~s^{-1}$ (Blue) and combining 
the resulting maps. These RGB $^{12}$CO(1-0) and 
$^{13}$CO(1-0) images are presented in Figures 
\ref{fig:12mom0color} and \ref{fig:13mom0color}, 
respectively, and provide broad kinematic information 
not discernible in the maps in Figures
\ref{fig:12mom0} and \ref{fig:13mom0}. 
As before, the intensity scale is 
in units of K $\rm km~s^{-1}$. 
The most impressive feature in both maps
is the large-scale, north-south,
velocity gradient across the length of the cloud,
However, it is not clear if the 
gradient is smooth or merely the result of overlapping clouds
along the line of sight.

Due to strong contributions from all three 
velocity ranges to the
$^{12}$CO(1-0) emission,
the $\sim 6\arcmin$ radius area around 
ONC appears white in Figure \ref{fig:12mom0color}.
This area coincides 
with the very active OMC-1 region, where a wide range 
of velocities results from the entrainment of the 
molecular gas by powerful outflows, especially those associated with the BN/KL objects
\citep{2017ApJ...837...60B}. 
The two areas of anomalously blue emission
seen in the otherwise red-velocity dominated
region just north of OMC-1 
can also be explained in light of previous
studies. The northernmost appears to be
associated with the Northern Ionization Front (NIF) 
\citep{Berne2014}, presumably where
the foreground molecular cloud material 
and the front of the Orion Nebula HII region interact.
Green and orange diagonal striations northwest 
of OMC-1 imply strong feedback effects  
from the outflow sources in OMC-1. 

The strong red emission observed
immediately south-southeast of the ONC is
consistent with the long-standing hypothesis that
this feature results from the interaction 
between the far side of the expanding Orion Nebula HII region 
and the molecular cloud 
\citep{1979ApJ...234L.207L,Berne2014}.
The irradiated pillar-like globules noted in 
Figure \ref{fig:12mom0}
to the south and southeast of the ONC,
appear as green and blue velocity features in
Figures \ref{fig:12mom0color} and \ref{fig:13mom0color}. 
They are thought to be primarily 
in the ``Dark Lane South Filament'' (DLSF)
on the near side of the cloud 
\citep{1998A&A...329.1097R,2011PASJ...63..105S,2013ApJ...774L..20S}.
Similar mixtures of blue and green
emission are seen in the L1641 cloud
and in the southern part of the ISF.
\citet{2012ApJ...746...25N} have 
proposed that a cloud-cloud collision
in the area triggered the formation 
of the filaments and initiated star 
formation in L1641-N cluster.

\begin{figure*}[htbp]
\epsscale{1.1}
\plotone{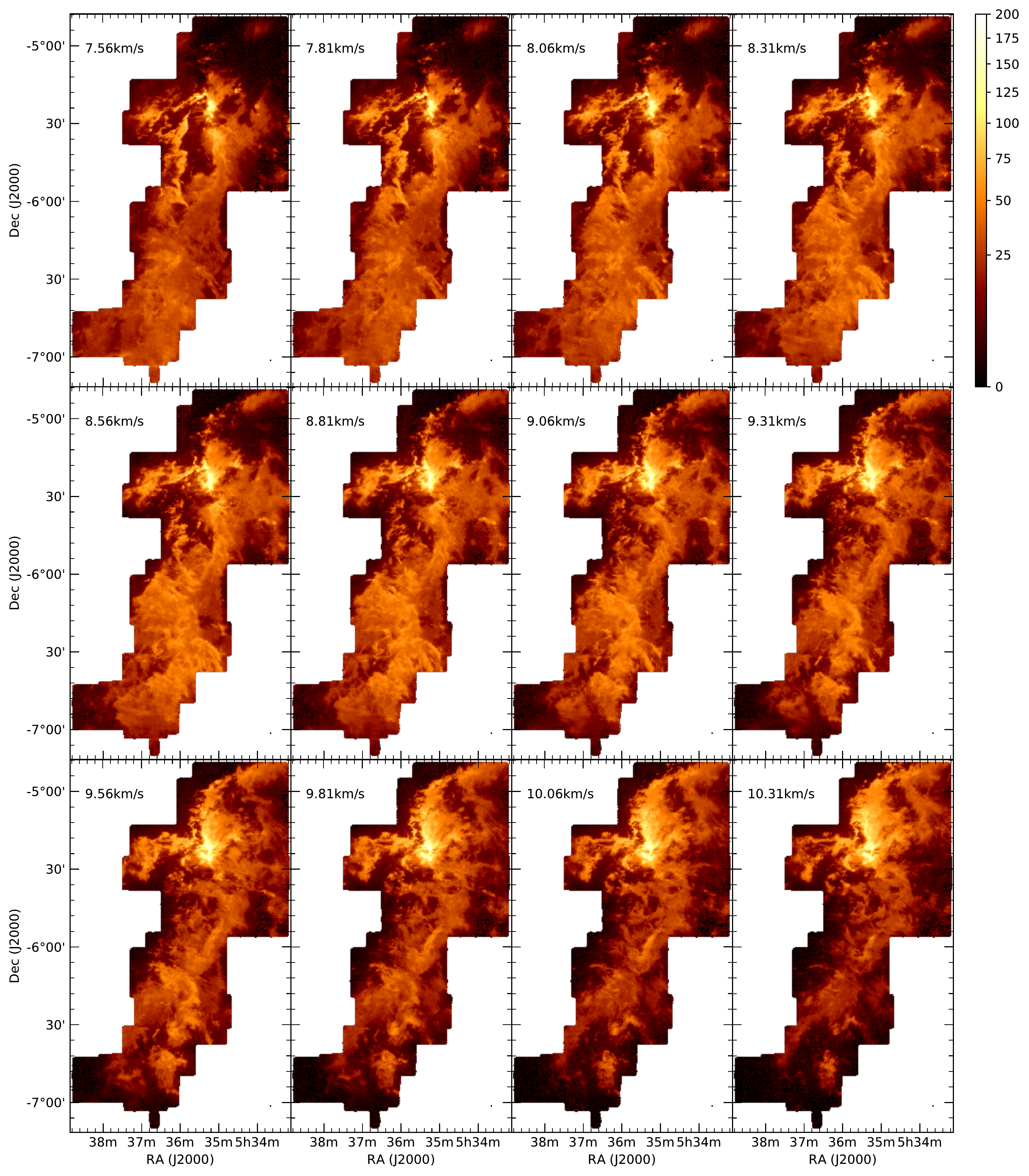}
\caption{
Channel maps for $^{12}$CO(1-0)
between 7.56 and 10.31 $\rm km~s^{-1}$.  
The color bar to the right is a square root 
scale in units of K. A channel map movie, 
including all channels,
is available online. Figure \ref{fig:chan12}, 
in the Appendix, shows all channel maps.
\label{fig:12chan}}
\end{figure*}

\begin{figure*}[htbp]
\epsscale{1.1}
\plotone{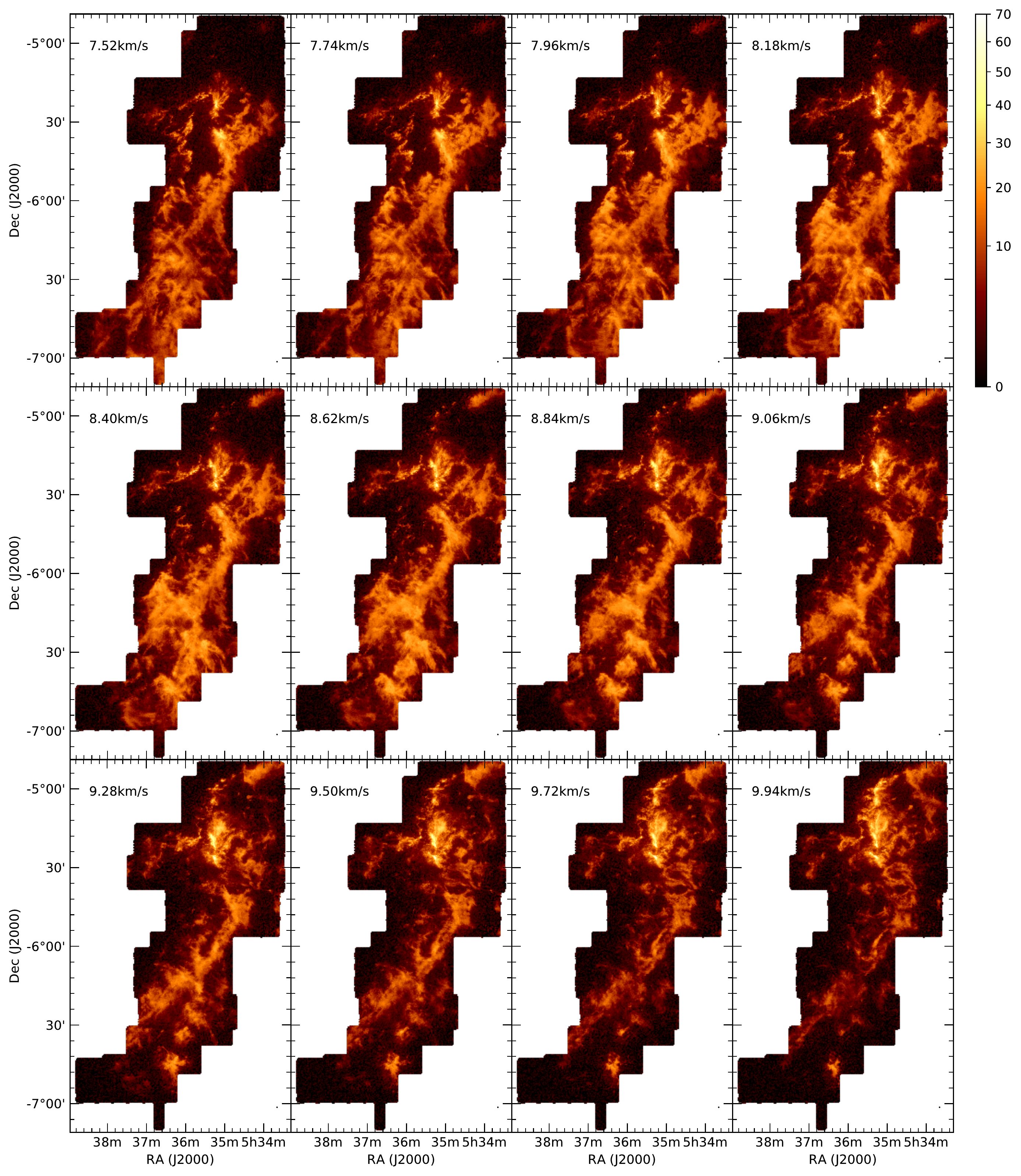}
\caption{
As Figure \ref{fig:12chan},
but for the $^{13}$CO(1-0) emission.
A channel map movie, including all channels,
is available online. Figure \ref{fig:chan13}, 
in the Appendix, shows all channel maps.
\label{fig:13chan}}
\end{figure*}

\begin{figure*}[htbp]
\epsscale{1.1}
\plotone{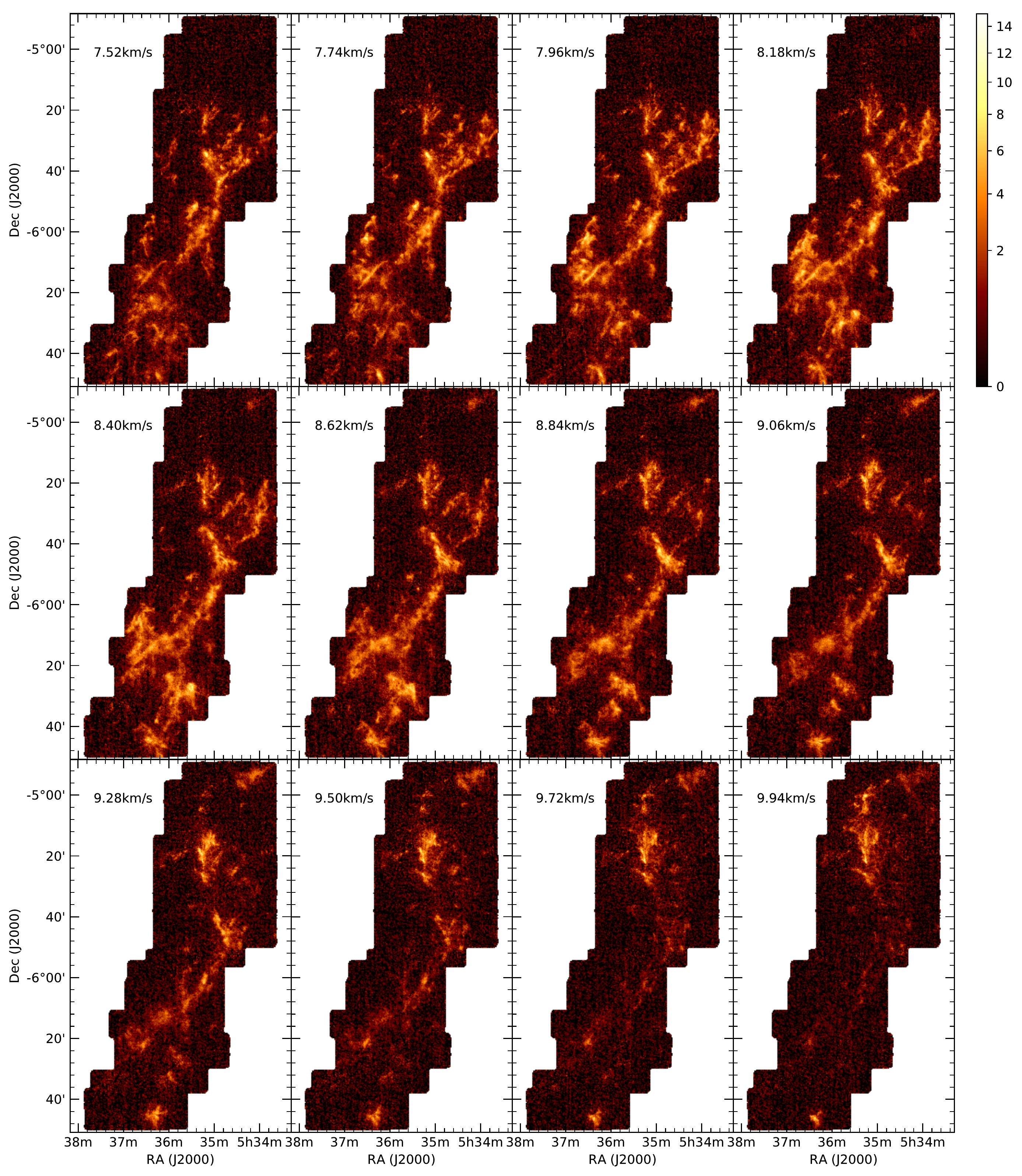}
\caption{
As Figure \ref{fig:12chan},
but for the C$^{18}$O(1-0) emission.
A channel map movie, including all channels,
is available online. Figure \ref{fig:chan18},
in the Appendix, shows all channel maps.
\label{fig:18chan}}
\end{figure*}

Velocity channel maps from our combined data cubes,
reveal the complex kinematics of Orion A in considerable detail.
Figures \ref{fig:12chan}, \ref{fig:13chan}, 
and \ref{fig:18chan}, display sample channel maps for
$^{12}$CO(1-0), $^{13}$CO(1-0), and C$^{18}$O(1-0), respectively. 
Movies showing channel maps over the full bandwidth for 
each cube are linked to the  online versions of these figures. 
Complete channel maps for all three cubes at full velocity resolution
are presented in appendix \ref{sec:chanmaps}.
Features clearly visible in these maps include the clumpy 
structure at various velocities in the L1641 region, 
the high-velocity gas from the Orion KL (OMC-1) explosive outflow 
and various filaments in the OMC-2/3 region.

\begin{figure*}[htbp]
\epsscale{0.33}
\plotone{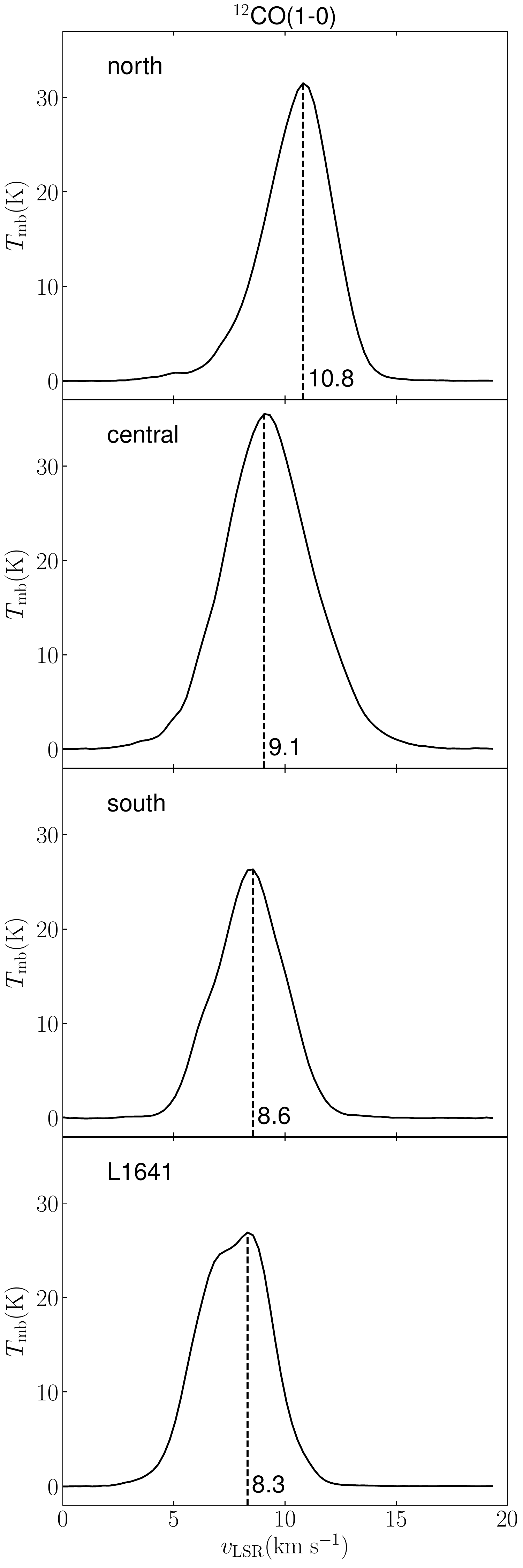}
\plotone{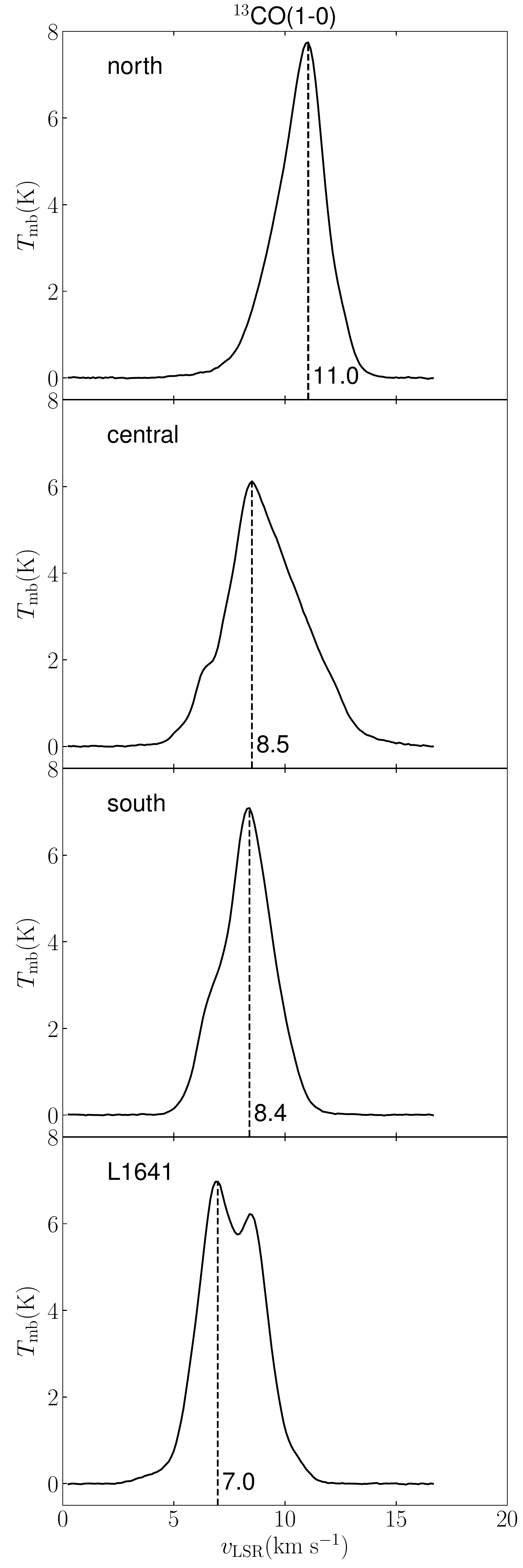}
\plotone{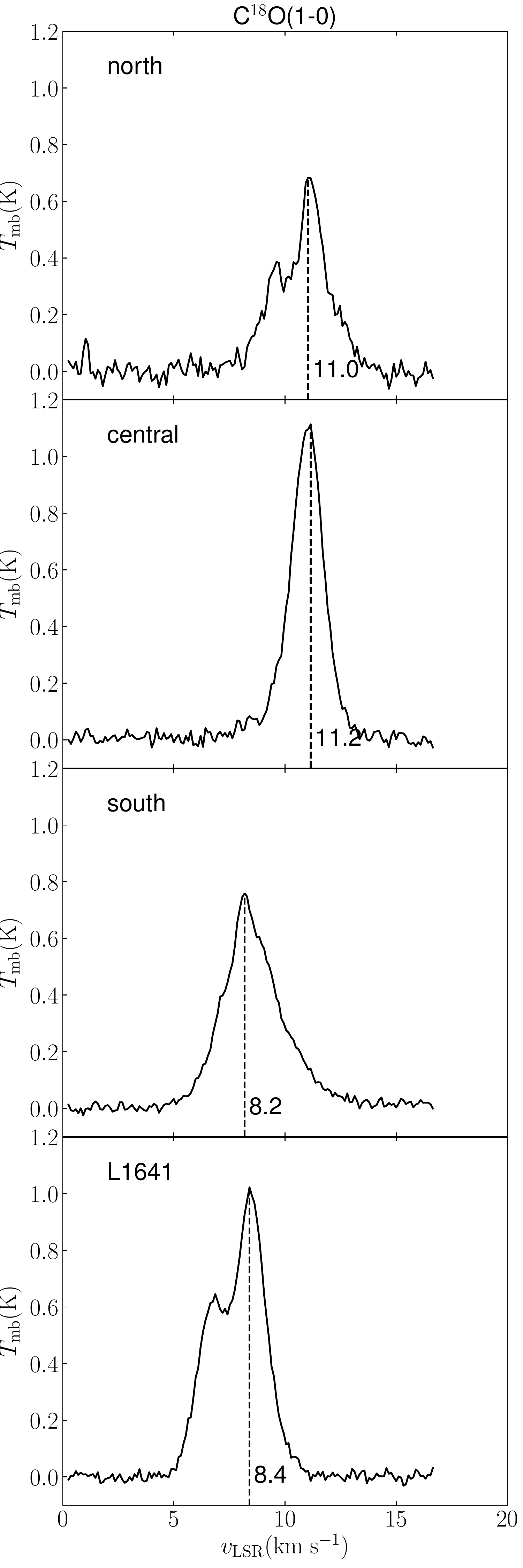}
\caption{Averaged profiles for $^{12}$CO(1-0) ($\it Left$), $^{13}$CO(1-0)
and ($\it Middle$), and C$^{18}$O(1-0) ($\it Right$)
in the four areas of the Orion A cloud shown
in Figure \ref{fig:coverage}.
The line profiles were constructed for
each separate region by averaging the 
intensities in each velocity channel.
Different intensity scales are required for
each molecular species but the
velocity scale is the same for all. 
Vertical dashed lines mark the velocity 
of peak emission in each case.
\label{fig:spec}}
\end{figure*}

The north-south velocity shift seen in the
RGB maps of Figures \ref{fig:12mom0color} 
and \ref{fig:13mom0color} can also be traced by considering 
the molecular line emission at various positions
across the Orion A cloud. 
Figure \ref{fig:spec} shows the averaged 
spectral line profiles for $^{12}$CO(1-0), 
$^{13}$CO(1-0), and C$^{18}$O(1-0)
for the four regions of the cloud defined 
in Figure \ref{fig:coverage}.
As expected, the velocity of the peak 
intensity for each CO species shifts 
from $\sim$ 11 $\rm km~s^{-1}$ to 7.5 $\rm km~s^{-1}$,
moving from north to south in the cloud, 
consistent with the large-scale
change in velocity seen in the RGB maps. 
The double-peaked lines seen in the
L1641 "region" have been noted
previously and gave rise to the 
cloud-cloud collision scenario
suggested by \citet{2012ApJ...746...25N}.

\begin{figure*}[htbp]
\epsscale{1.1}
\plotone{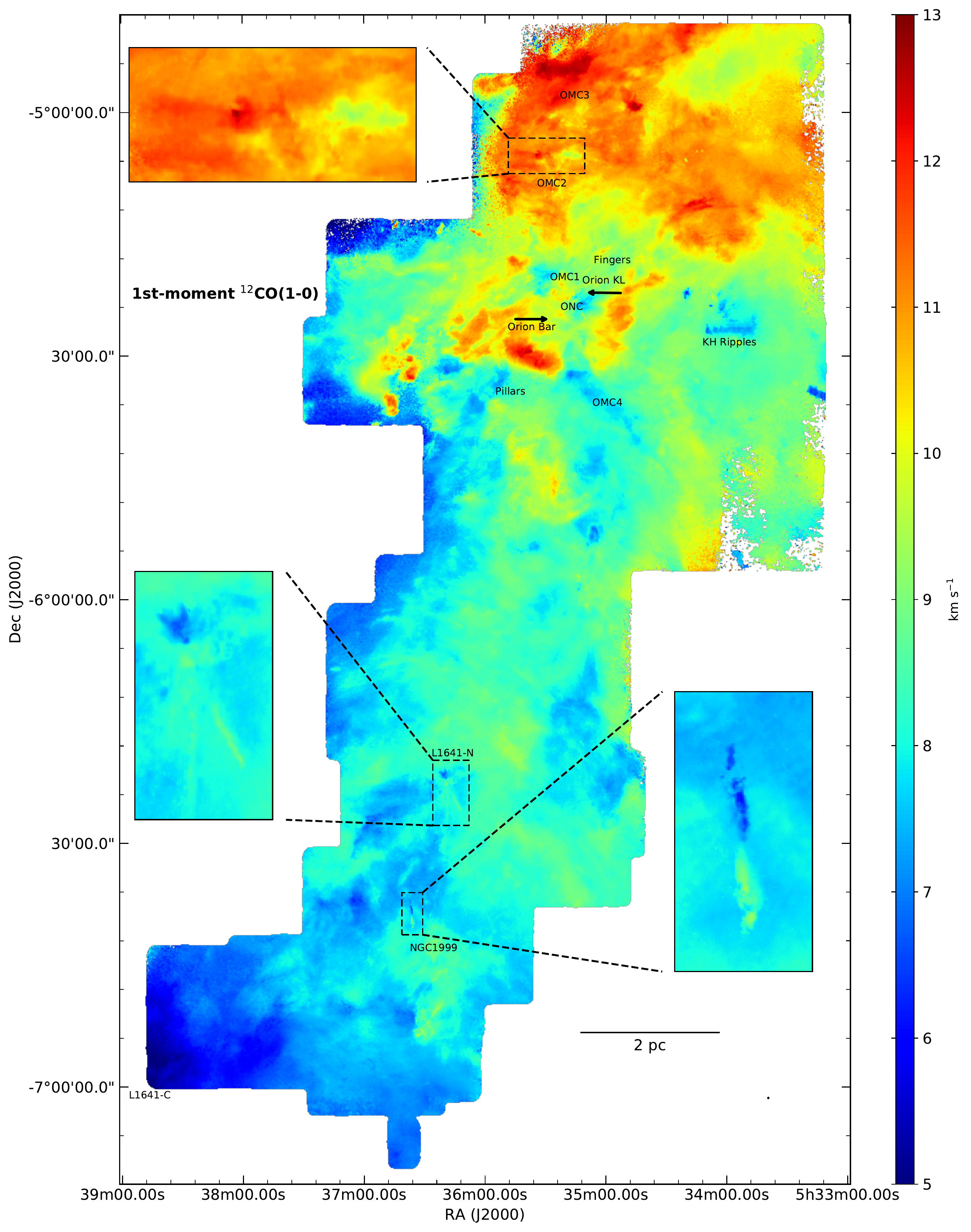}
\caption{
First moment map showing the variation of peak $^{12}$CO(1-0) velocities
between 2.5 and 15 $\rm km~s^{-1}$
as a function of position in the Orion A cloud.  
Only emission above 5$\sigma$ is used to produce the map.
The color bar to the right is a linear 
scale in units of $\rm km~s^{-1}$.
Insets show zoom-in views of three interesting outflows.
\label{fig:12mom1}}
\end{figure*}

\begin{figure*}[htbp]
\epsscale{1.1}
\plotone{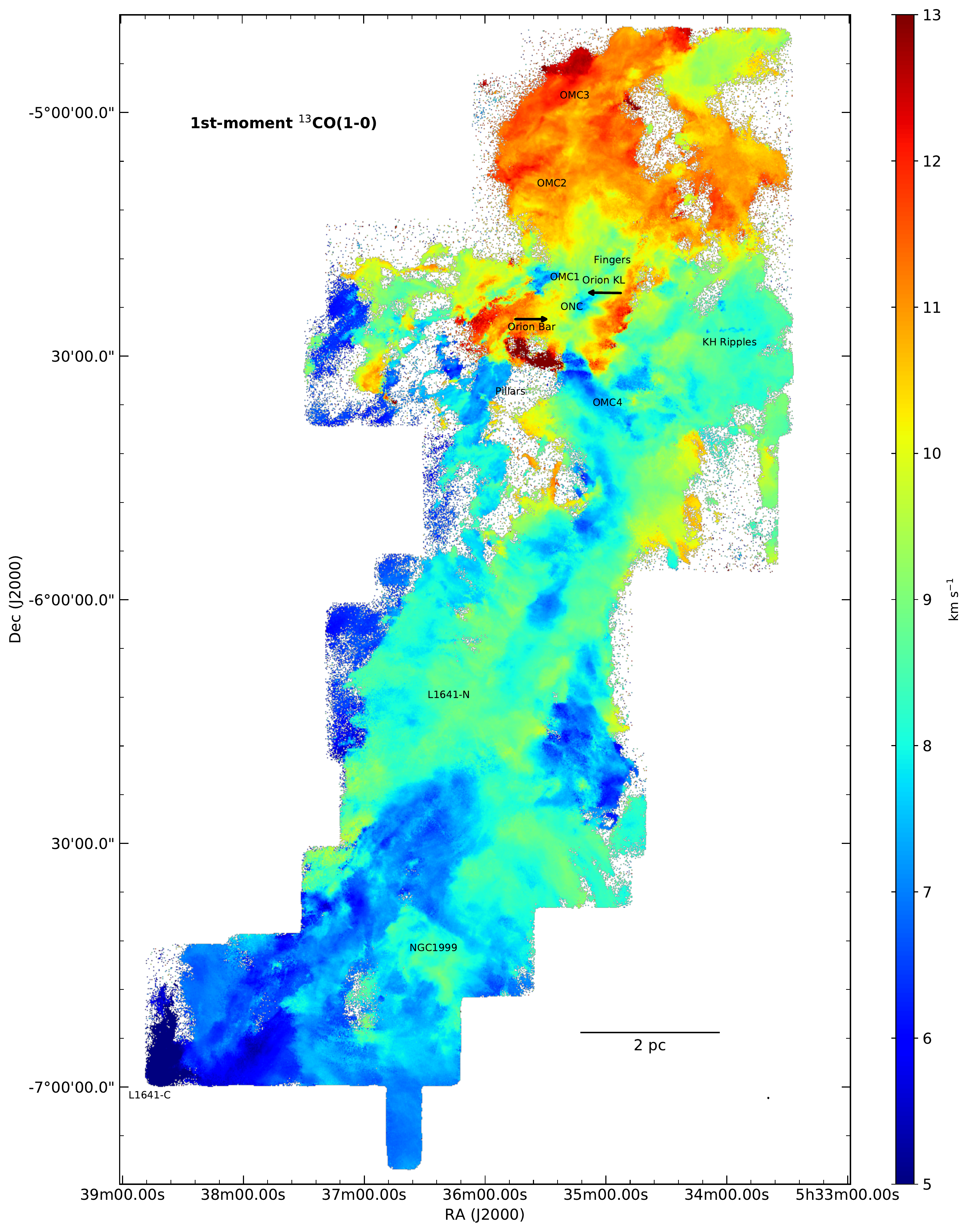}
\caption{
As Figure \ref{fig:12mom1},
but for the $^{13}$CO(1-0) emission.
\label{fig:13mom1}}
\end{figure*}

\begin{figure*}[htbp]
\epsscale{1.1}
\plotone{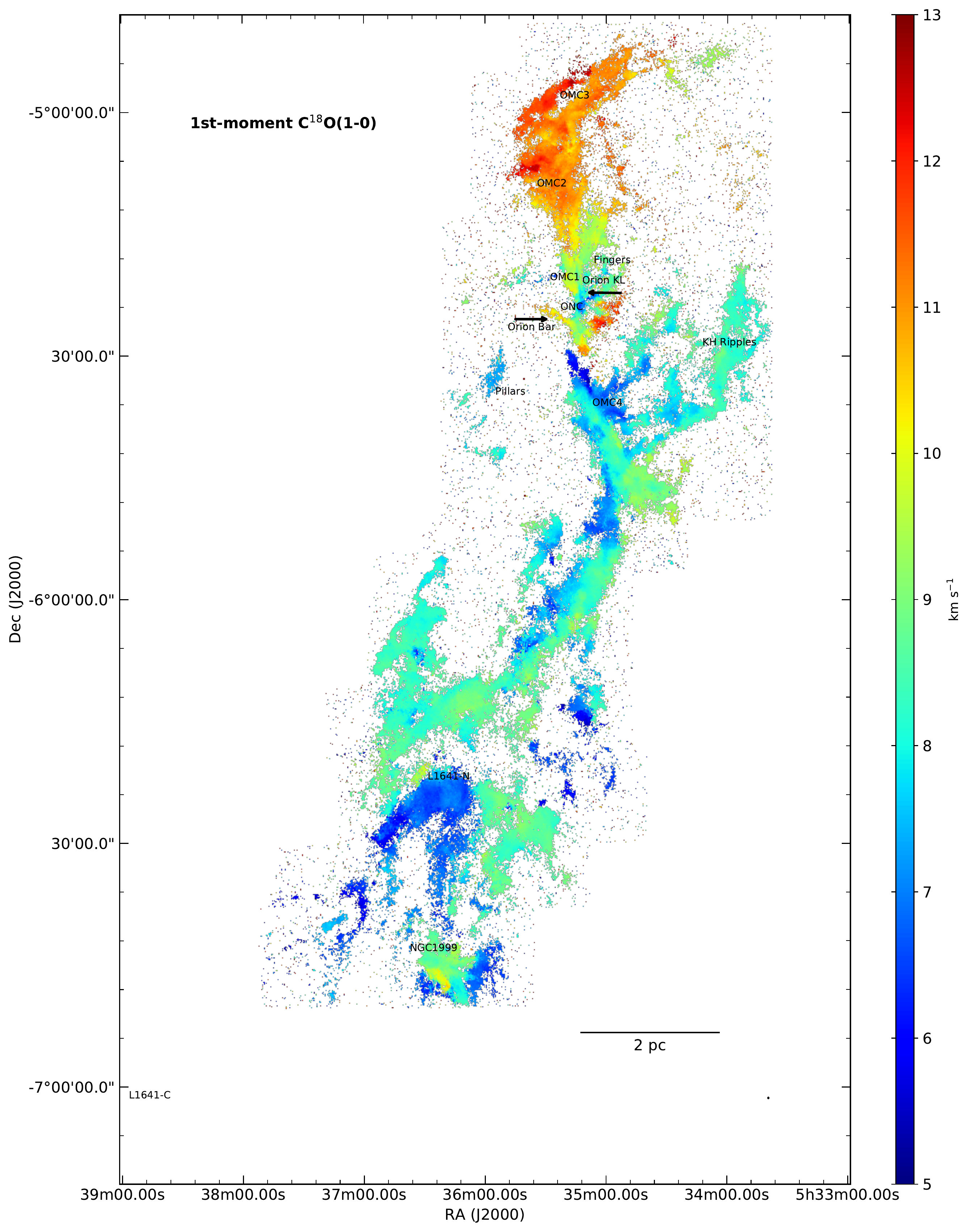}
\caption{
As Figure \ref{fig:12mom1},
but for the C$^{18}$O(1-0) emission.
\label{fig:18mom1}}
\end{figure*}

The overall north-to-south shift in velocity 
can also be seen in 
our 1st-moment maps for $^{12}$CO(1-0), $^{13}$CO(1-0),
and C$^{18}$O(1-0), Figures \ref{fig:12mom1},
\ref{fig:13mom1}, and \ref{fig:18mom1},
respectively. Again, the northern part of
the cloud is preferentially redder, indicating higher 
velocities, while the southern part is, for
the most part, blue and at lower velocities. 
The origin of the gradient 
has been discussed at length in the literature 
but is not yet fully understood.
To date, explanations put forward include
rotation, filament expansion, and global collapse 
\citep[][and references therein]{2008hsf1.book..621A}.
The high resolution, high sensitivity CARMA-NRO Orion 
dataset presented here offers a new opportunity
to  test every hypothesis in detail over the 
requisite extended scales.

In addition, these 1st-moment maps present,
in much  more detail, the variation in velocity
with position in the cloud, over the entire
velocity integration range,
2.5 to 15 $\rm km~s^{-1}$. Like the maps 
of gas distribution, they show interesting 
features, often noted by earlier focused studies, in a 
larger-scale context,
Among these, is a distinctive elongated blue field 
to the west of OMC-1 in Figure \ref{fig:12mom1}. This coincides with the
location of the Kelvin-Helmholtz ``ripples''
that are thought to be in the foreground of the ONC \citep{2010Natur.466..947B}. Likewise,
in Figure \ref{fig:18mom1} there appears to be a 
velocity gradient along the short axis of the 
filament below OMC-4, similar to that
seen in simulations of C$^{18}$O(1-0) emission from 
a filament in a turbulent cloud by
\citet{2016MNRAS.455.3640S}. These authors attributed
the gradient to the 
filament moving across the cloud as a
coherent velocity front, corresponding to 
one of the large scale modes in a turbulent 
cascade. Given the results of our delta variance analysis,
it is tempting to suggest that
a similar mechanism may be at work here. 
More importantly, this highlights yet again, the potential 
impact of our new survey on the understanding of 
large-scale processes in molecular clouds.

Interesting features in Figure \ref{fig:12mom1} 
include several prominent bipolar outflows. 
They are noticeable because the velocities of
the outflowing molecular gas differ significantly
from the systemic cloud velocity in their neighborhoods,
increasing the contrast with the cloud.
The effect is most easily seen in the Figure 
\ref{fig:12mom1} insets. 
In the southern part of the cloud, elongated 
deep blue outflow lobes contrast with  
cyan/light green regions. In the north, 
extended red/yellow lobes contrast with light/green structures. 
Identified examples of bipolar outflows in the map include 
V380 Ori-N \citep{2000MNRAS.318..952D}, 
L1641-N \citep{2007AJ....133.1307S}, and
OMC-2/3 MMS 9  \citep[e.g.,][]{2008ApJ...688..344T}. 
There are certainly many more outflows in the region
\citep[see, e.g.,][]{2003ApJ...591.1025W,2008ApJ...688..344T,2012ApJ...746...25N}. 
However, if these are low-intensity emission, 
with velocities similar to the systemic value, or
are small or unipolar, they will be   
difficult to detect in the first moment map.

\begin{figure*}[htbp]
\epsscale{1.1}
\plotone{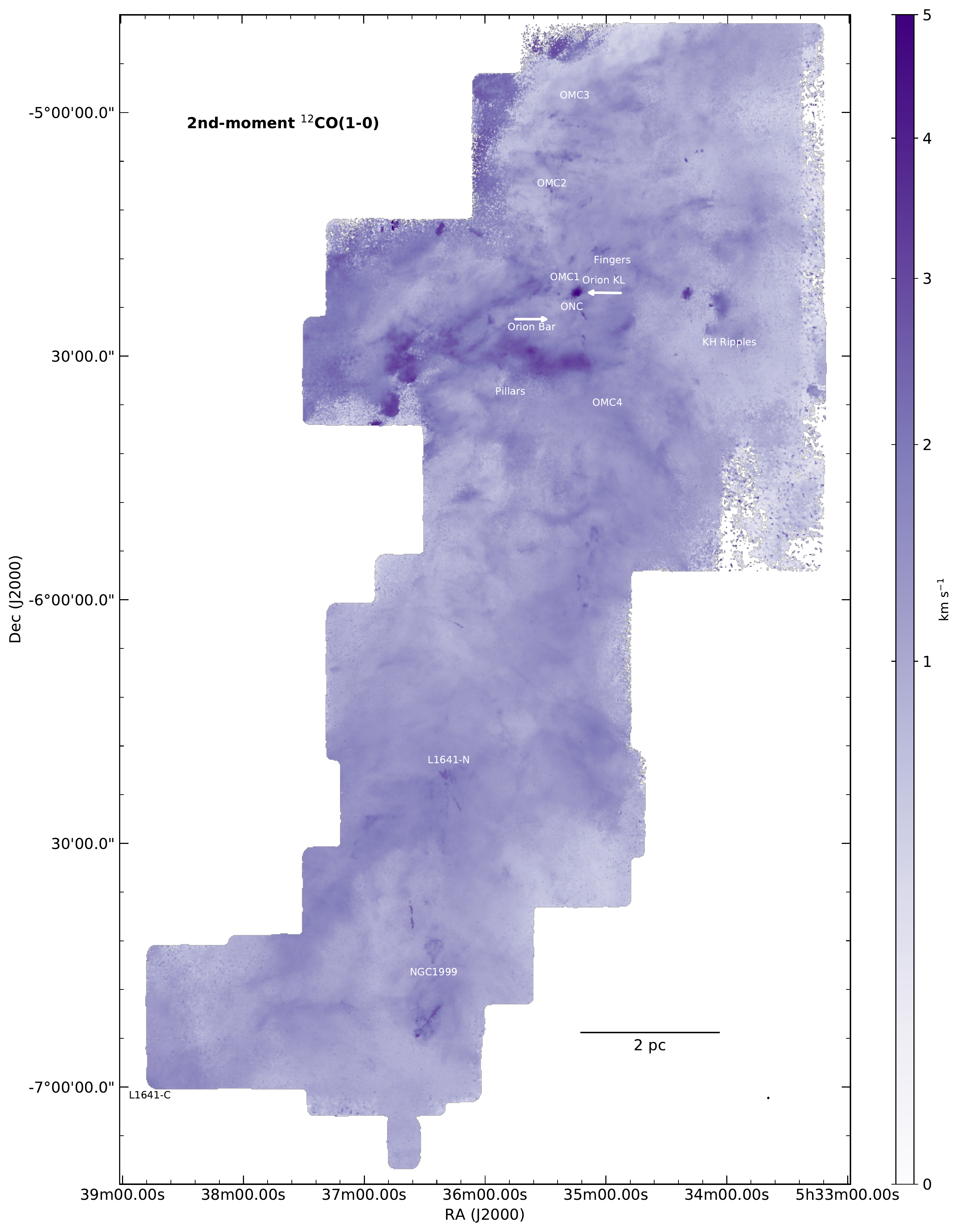}
\caption{
Second moment map of the $^{12}$CO(1-0) emission
over the velocity range 2.5 to 15 $\rm km~s^{-1}$ 
showing the variation of the velocity dispersion of $^{12}$CO.
Only emission above 5$\sigma$ is used to produce the map.
The color bar has a square root scale with 
units of $\rm km~s^{-1}$.
\label{fig:12mom2}}
\end{figure*}

\begin{figure*}[htbp]
\epsscale{1.1}
\plotone{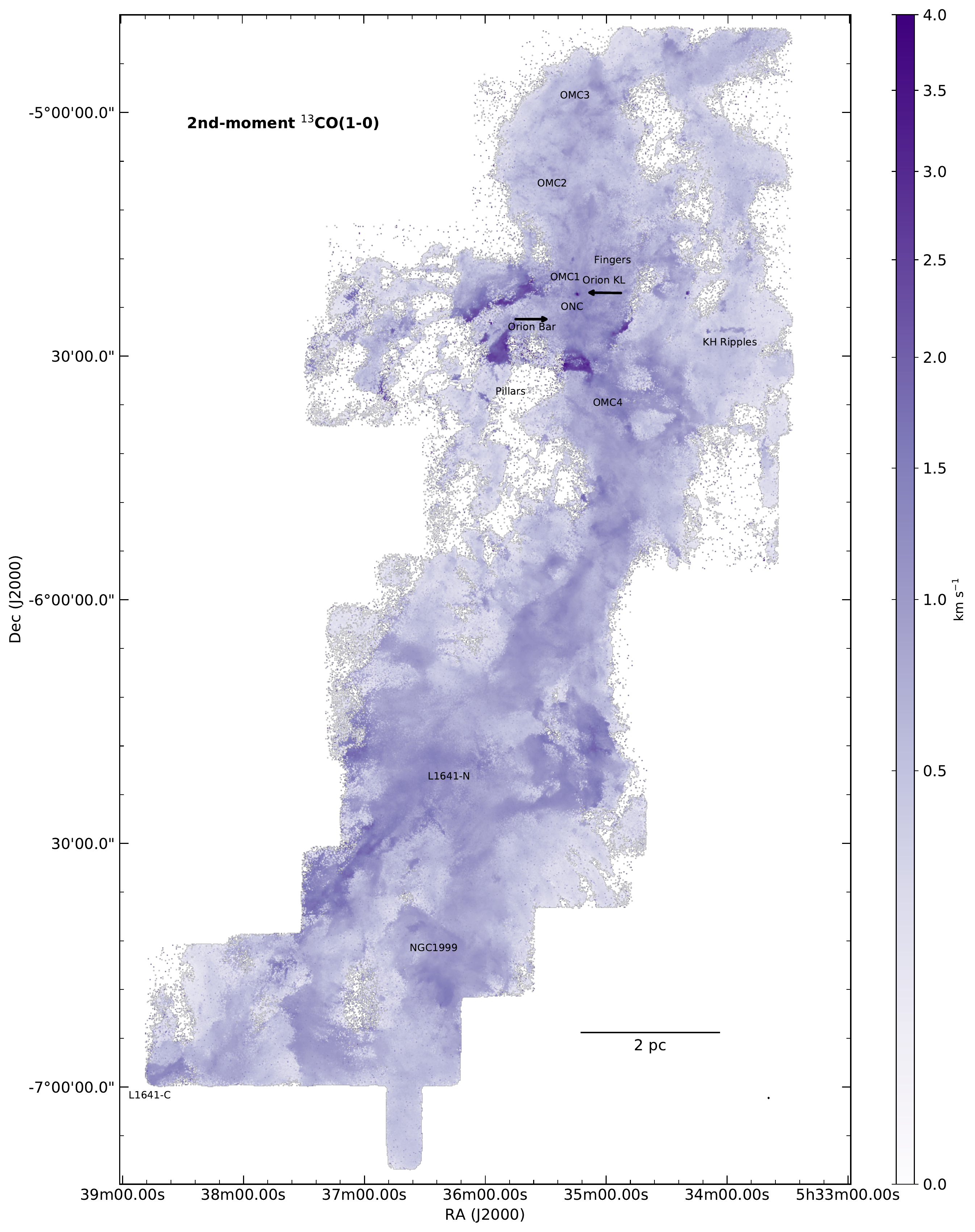}
\caption{
As Figure \ref{fig:12mom2},
but for $^{13}$CO(1-0).
\label{fig:13mom2}}
\end{figure*}

\begin{figure*}[htbp]
\epsscale{1.1}
\plotone{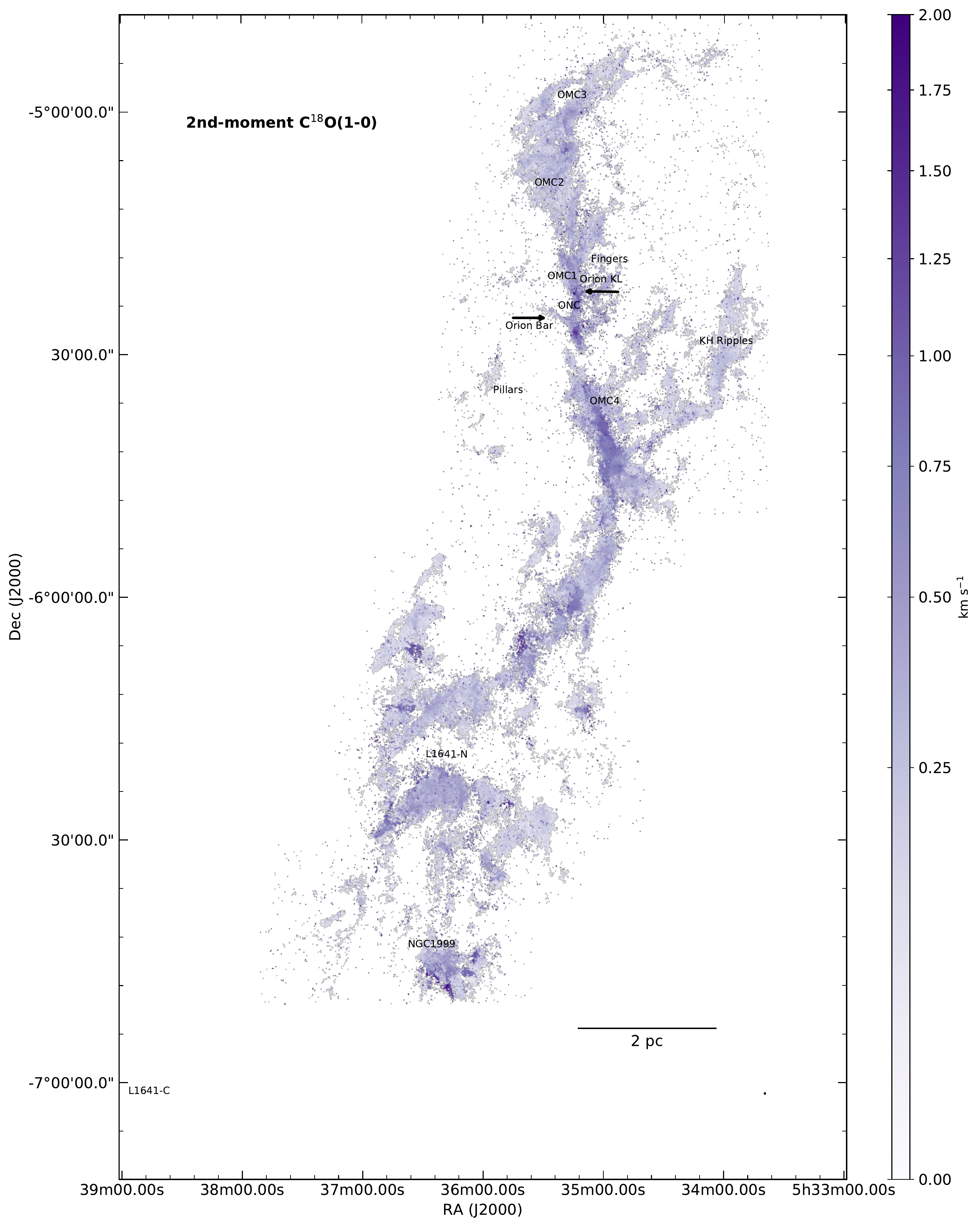}
\caption{
As Figure \ref{fig:12mom2},
but for the C$^{18}$O(1-0).
\label{fig:18mom2}}
\end{figure*}

Signposts of feedback can also be detected in
maps that display the variation of velocity 
dispersion across the Orion A cloud. 
These are provided in second moment maps of
$^{12}$CO, $^{13}$CO, and C$^{18}$O in
Figures \ref{fig:12mom2}, \ref{fig:13mom2}, 
and \ref{fig:18mom2}, respectively.
For the most part, all three maps show a smooth distribution, 
with dispersions mostly between 0.5 to 
1.5 $\rm km~s^{-1}$ but occasionally  
reaching 2 $\rm km~s^{-1}$). 
The highest values coincide with the Orion KL region  
and are likely attributable to the high-velocity explosive 
outflow 
\citep{2017ApJ...837...60B}.
Significantly high velocity dispersions 
are also noticeable in 
the region around the ONC characterized as the ``shell
around Orion-KL'' \citep{2011PASJ...63..105S}, 
and may be due to feedback effects from ONC region and
overlapping line-of-sight velocity 
components.
In Figure \ref{fig:12mom2},
high velocity dispersions are also associated with 
molecular outflows in NGC 1999, L1641-N, and 
OMC-2/3.

\begin{figure*}[htbp]
\epsscale{1.}
\plotone{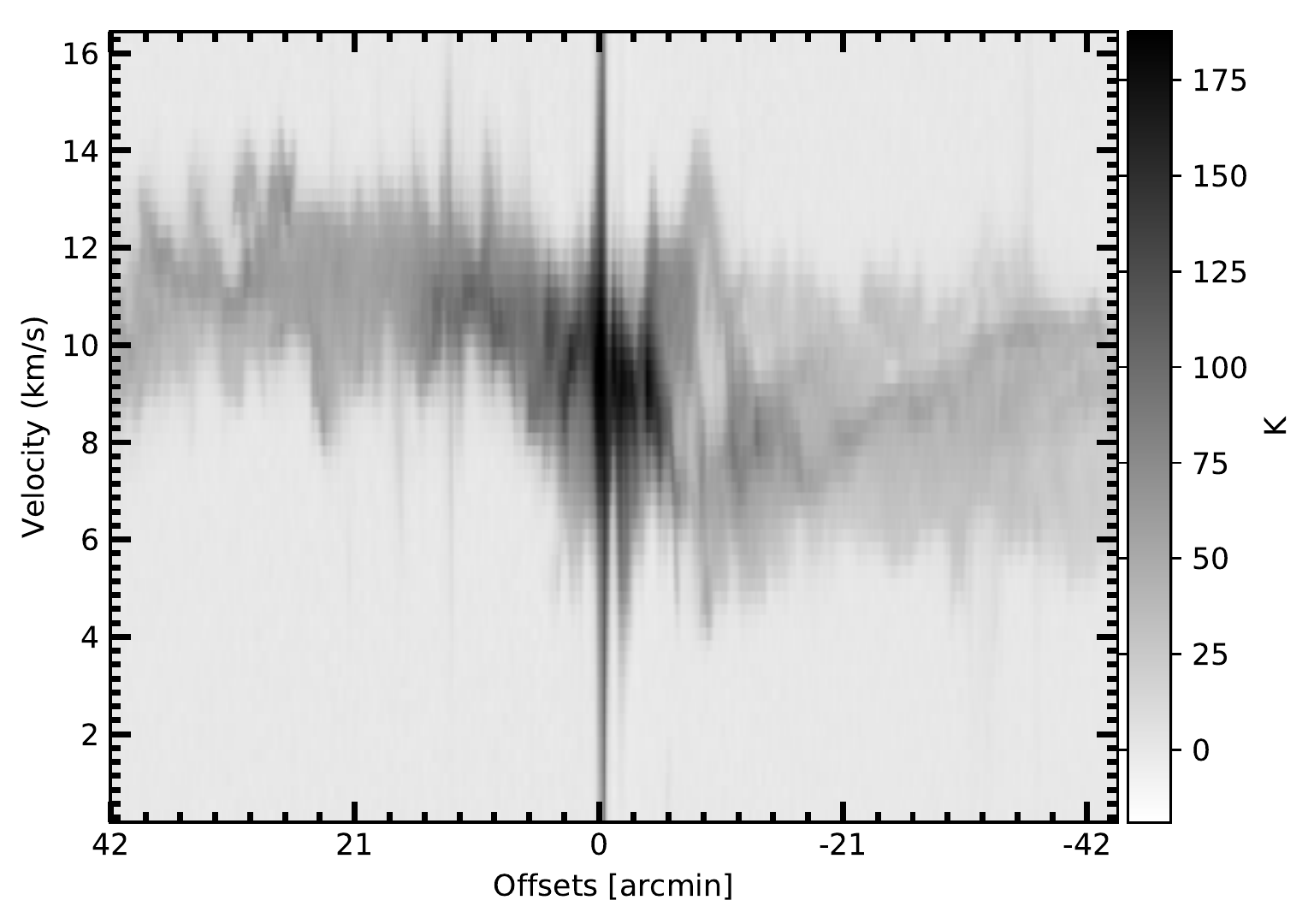}
\caption{
Position-velocity diagram 
of the $^{12}$CO(1-0) emission 
along the ISF (see Figure \ref{fig:12mom0}).
The reference (offset = 0) position coincides with  
the position of Orion KL, with positive offsets 
to the north. The gray-scale bar is linear 
and in units of main-beam temperature (K).
\label{fig:pv12}}
\end{figure*}

\begin{figure*}[htbp]
\epsscale{1.}
\plotone{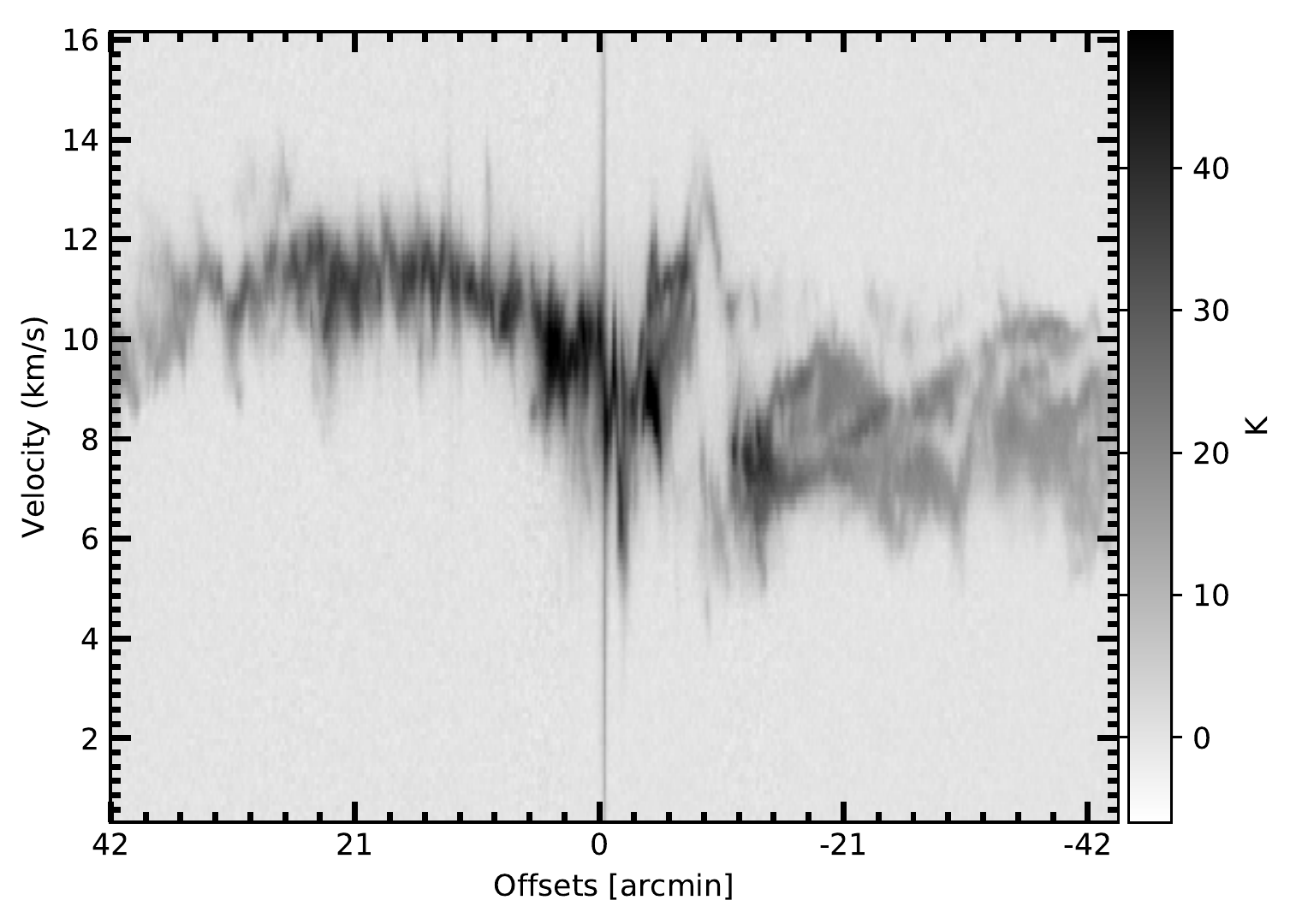}
\caption{
As Figure \ref{fig:pv12},
but for $^{13}$CO(1-0).
\label{fig:pv13}}
\end{figure*}

\begin{figure*}[htbp]
\epsscale{1.}
\plotone{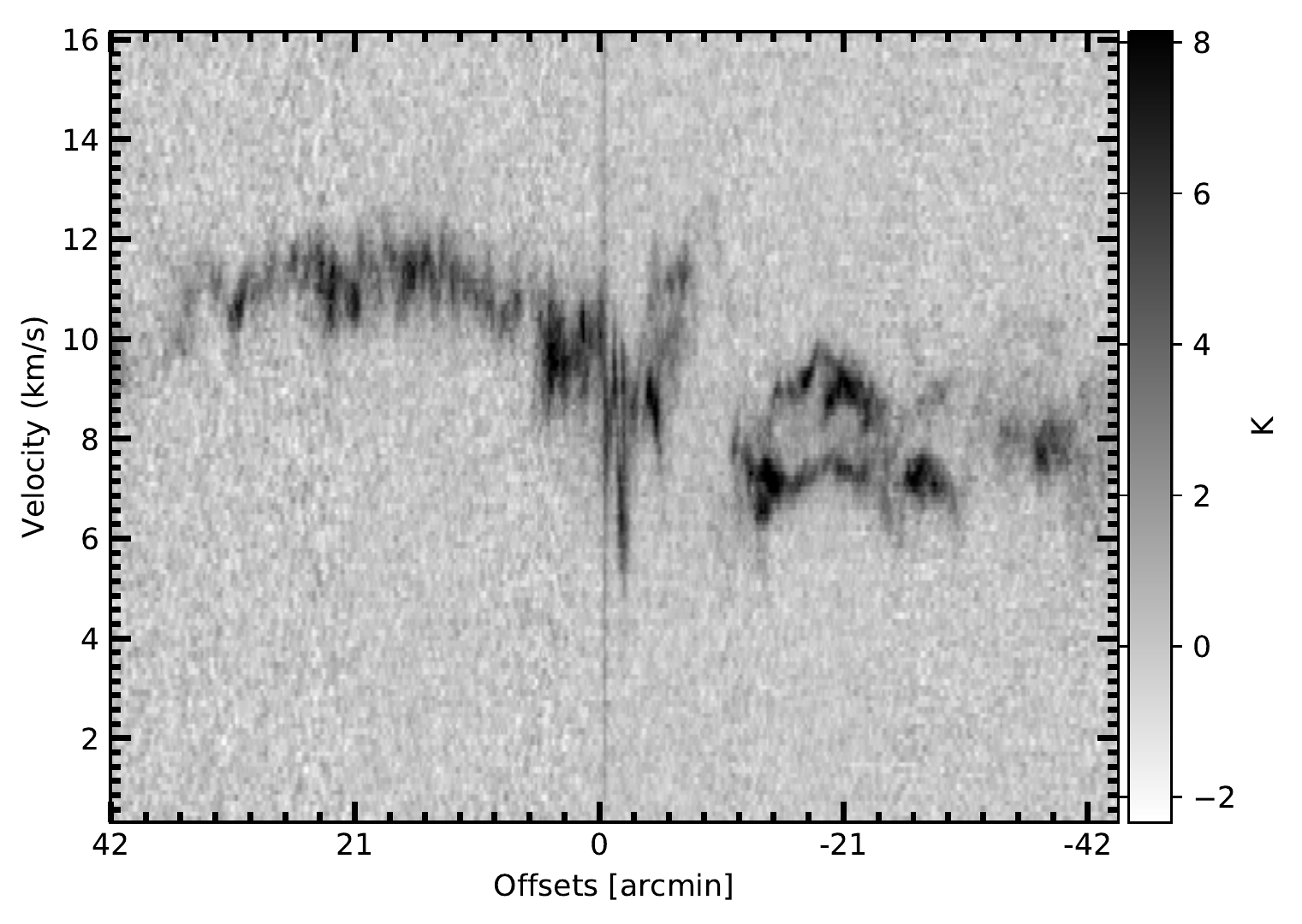}
\caption{
As Figure \ref{fig:pv12},
but for the C$^{18}$O(1-0).
\label{fig:pv18}}
\end{figure*}

Another way to examine how gas kinematics change with location in the cloud relies on position-velocity (PV) diagrams for  the regions of interest. 
As an example, in Figures \ref{fig:pv12}, \ref{fig:pv13}, and \ref{fig:pv18}, we show the $^{12}$CO, $^{13}$CO, and C$^{18}$O PV diagrams along the ISF. The blue curve in Figure \ref{fig:12mom0} defines the PV cut.
A large blue dot at the position of the Orion KL represents the reference position 
(RA$_{\rm J2000}$ = $\rm 5^h35^m13\fs0$, DEC$_{\rm J2000}$ = 
$−5\arcdeg22\arcmin05\farcs0$). Positive offsets are toward the north. 
We extract a beam-averaged spectrum every 3\arcsec\ along the PV cut.
The velocity resolution in the PV diagrams
for each cube are as given in Table \ref{tab:sensitivity}.

The broad velocity spread at the 
the reference position is prominent in
all three PV diagrams, but especially in 
Figure \ref{fig:pv12}, and reflects the high-velocities of the explosive outflow known to be present in the Orion KL region \citep{2017ApJ...837...60B}. 
Other high-velocity spikes 
can be discerned in Figure \ref{fig:pv12}, 
at offsets 16\arcmin, 12\arcmin, 
and -37\arcmin, for example, and are likely due to bipolar outflows. 
A bubble-like structure at offset -10\arcmin\ in
Figures \ref{fig:pv13} and \ref{fig:pv18}
is probably associated with the ``shell around 
Orion-KL'' mentioned earlier.
The overall sinusoidal shape present in our PV diagrams is very intriguing.  \citet{Stutz2016} and \citet{2018MNRAS.475..121S} proposed the slingshot model in which the gas is oscillating, ejecting protostars, in an auto-destructive cycle that ultimately results in cluster formation.  The observed morphology in our PV diagrams is strongly reminiscent of such a wave-like motion.  However, the exact nature of this wave remains unclear.  Our observed signature is inconsistent with a propagating wave because this would require that the maximum displacement corresponds to the minimum velocity, in  contradiction with Figures \ref{fig:pv12},\ref{fig:pv13},\ref{fig:pv18}.  These figures show that the maximum displacement coincides with the position of maximum velocity, most clearly seen in the OMC-2/3 regions, located near offset 20\arcmin.  Waves of other types (such as standing or torsional) may better explain the data. Further analysis is needed for the types of instabilities that may give rise to our observed PV diagram morphology \citep[e.g.,][]{2018MNRAS.475..121S}.
This is one example where the data presented here could be 
use to constrain models of different kinds of processes in
molecular cloud, and underlines the vast
reservoir of kinematic information
that becomes available with the CARMA-NRO survey.
Overall, the results provide an unprecedented 
opportunity for increasing our understanding of
energetic phenomena in star forming clouds. 

\section{Discussion}

\subsection{Gas Excitation Temperature}\label{subsec:tex}

\begin{figure*}[htbp]
\epsscale{1.1}
\plotone{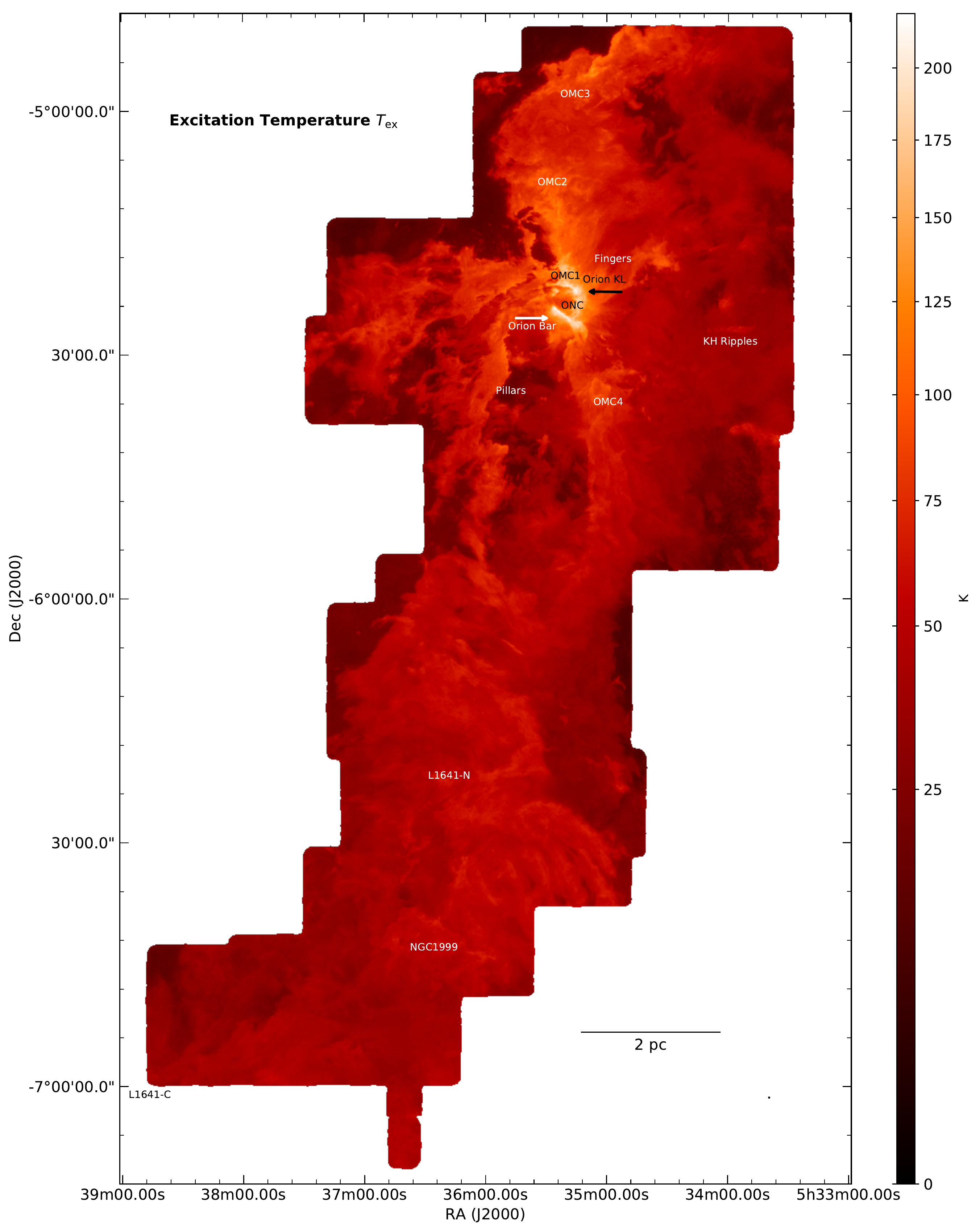}
\caption{
Excitation temperature $T_{\rm ex}$ map derived
from $^{12}$CO(1-0) peak intensity.
The color bar is in square root scale with 
units of K.
\label{fig:tex12}}
\end{figure*}

Measurements of the excitation temperature and
column density of the molecular gas across the 
Orion A cloud are critical to interpreting 
the results presented thus far. Figure \ref{fig:tex12}
shows the variation of excitation temperature, 
$T_{\rm ex}$, across the mapped area of the Orion cloud.
Values of $T_{\rm ex}$ were determined from the peak 
intensities of the $^{12}$CO(1-0) line in velocity space, 
$T_{\rm max}(^{12}{\rm CO})$, 
assuming the line is optically thick.  
Following \citet{2008ApJ...679..481P}, we use
\begin{equation}\label{eq:tex}
T_{\rm ex} = \frac{5.5~{\rm K}}{{\rm ln} \{ 1+5.5~{\rm K}/[T_{\rm max}(^{12}{\rm CO})+0.82~{\rm K}] \} }.
\end{equation}
Maximum and minimum derived values for $T_{\rm ex}$ 
in the map are 220 K and 2.8 K, respectively,
with a mean of 43.6 K and  a median value 
of 41.2 K. As a check, we compared our derived excitation
temperatures with those derived from the IRAM 30m $^{12}$CO(2-1)
map by \citet{Berne2014} that
covers the region from OMC-3 in the north 
to just south of OMC-4. The combined CARMA+NRO45 map was smoothed
to the $\sim 11\arcsec$ resolution of the IRAM 
30m map and $T_{\rm ex}$ estimated from peak intensity of the
line at each position. Within the inner quarter of the IRAM 
30m map, with the best signal to noise ratio, most values 
coincide within 35\%.

The excitation temperature distribution in 
Figure \ref{fig:tex12} shows a number of trends.
At the northern boundary of the map, in effect the
northern edge of OMC-3, temperatures
range from 80 to 110 K. It is likely that these relatively 
high temperatures result from the 
interaction of ionizing photons from the 
NGC 1973/1975/1977 H II region beyond the map boundary. 
Moving south towards OMC-2 along the ISF, 
temperatures first decrease ($T_{\rm ex} \sim 60$ K)
but rise again to the west of
M43 H II region. The highest temperatures  
are in the OMC-1/Bar region and are usually
around 170 K, but can reach 220 K. These values are consistent with
estimates of the gas kinetic temperature in the 
OMC-1 and Bar region derived from para-H$_2$CO line ratios 
\citep{2018A&A...609A..16T}, and also with
excitation temperatures derived from observations of CO(6-5)
\citep{2012A&A...538A..12P} and CO(9-8) 
\citep{2002A&A...394..271K}.

Southeast of the ONC, in the Pillars region,
$T_{\rm ex}$ always increases towards the tips of these structures
closer to the ONC, consistent with heating 
by the UV-radiation from its high-mass stars.
South of the ONC, temperatures continue to decrease 
along the ISF as the distance 
from the new high-mass stars increases, reaching about 50 K 
at the southern edge. Further south, average temperatures 
remain relatively stable. For example, in L1641, values of 30 to 40 K
are seen in the eastern region, and 40 to 50 K 
in the west. These are relatively high
temperatures compared to low-mass star forming regions
such as  Perseus and Taurus, where $T_{\rm ex} \sim 10-20$ K 
\citep{2008ApJ...679..481P,2010ApJ...721..686P}.
The enhancement is probably the result of CO tracing the warm 
surface of the cloud that is heated by the many 
relatively nearby  OB stars. 

\subsection{Gas Column Density}\label{subsec:coldens}

\begin{figure*}[htbp]
\epsscale{1.1}
\plotone{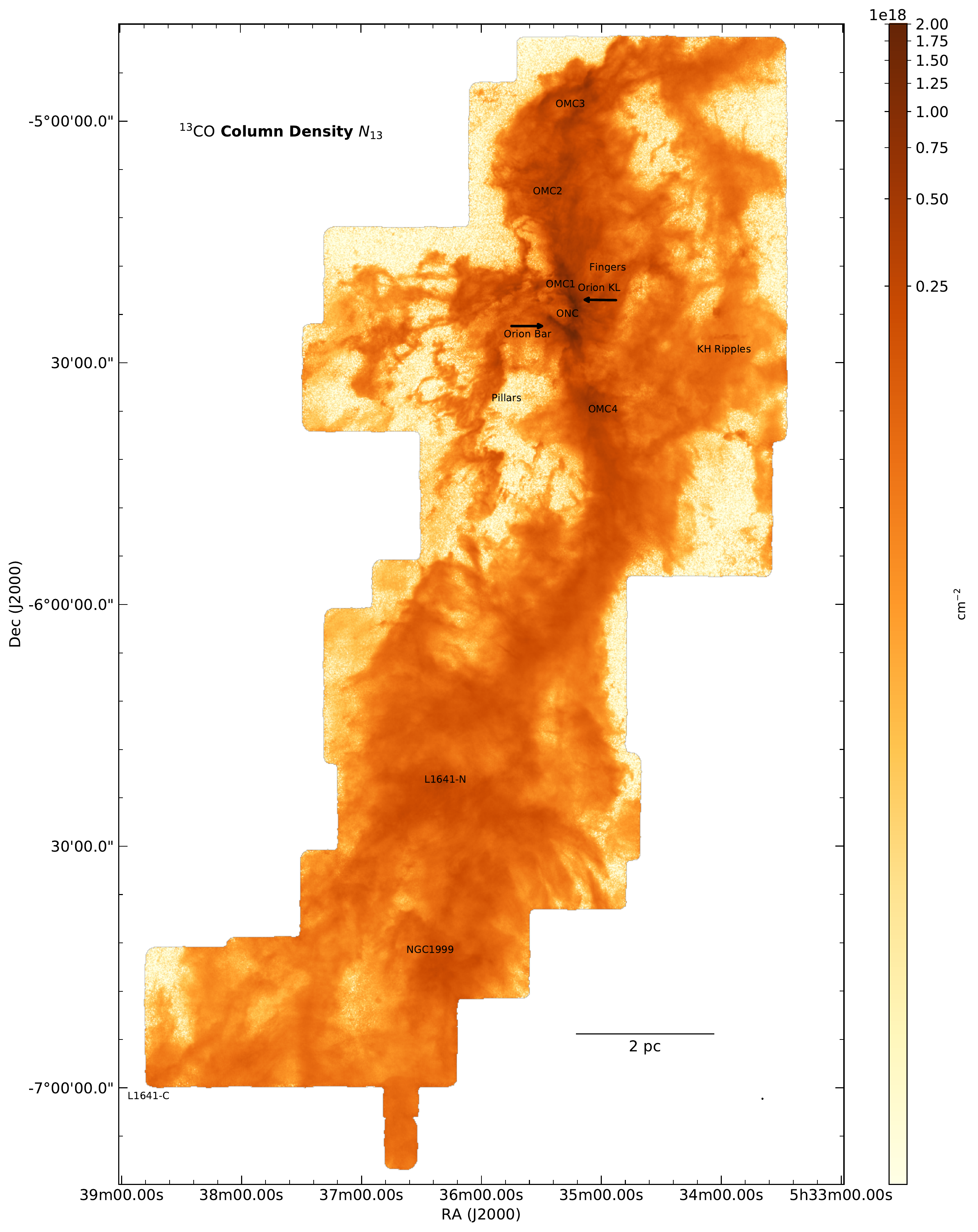}
\caption{
A map of the $^{13}$CO column density, $N_{13}$,
for the Orion A cloud. Units are 10$^{18}$ cm$^{-2}$.
\label{fig:coldens13}}
\end{figure*}

The column density of gas, $N_{13}$, was determined from  
our $^{13}$CO(1-0) data, assuming the excitation
temperatures derived for $^{12}$CO(1-0).
Following \citet{2008ApJ...679..481P},
$^{13}$CO optical depths, $\tau_{13}$, 
were calculated at each pixel. 
$^{13}$CO column densities, $N_{13}$,
were then calculated as described in 
\citet{1997ApJ...476..781B}, for those pixels
where the SNR exceeded 3. 
If the regions probed by $^{13}$CO(1-0) 
emission are cooler than the surface areas 
probed by $^{12}$CO, the absolute column densities
of $^{13}$CO will be underestimated. However, 
the overall variations across the map will be 
relatively unaffected. For most pixels, 
the emission is optically thin. Only about 
0.6\% of the pixels have $\tau_{13} > 1$ and 
these are mostly in the OMC-2/3 and L1641-N regions.

The resulting column density map is shown 
in Figure \ref{fig:coldens13}. Values of $N_{13}$ 
range from 2.1$\times$10$^{14}$ cm$^{-2}$ to
3.2$\times$10$^{18}$ cm$^{-2}$. The mean is
6.5$\times$10$^{16}$ cm$^{-2}$
and the median is 3.9$\times$10$^{16}$ cm$^{-2}$. 
The highest values of $N_{13}$ are seen in the OMC-1 
and Bar regions, with values of approximately  
$0.8 - 3 \times$10$^{18}$ cm$^{-2}$. Relatively 
high values of the column density,
$2-5\times$10$^{17}$ cm$^{-2}$, are also evident 
in the OMC-2 and OMC-3 regions of active star formation,
and column densities of $1-3\times$10$^{17}$ cm$^{-2}$ 
characterize the ISF south of the ONC. The L1641 region, 
south of -6\arcdeg10\arcmin, show relatively low
 column densities of about  10$^{17}$ cm$^{-2}$, 
and there are noticeable increases to 
$2-3 \times$10$^{17}$ cm$^{-2}$ in the L1641-N 
and NGC 1999 clusters. 

\subsection{The $^{12}$CO(1-0)/$^{13}$CO(1-0) line ratio} 

\begin{figure*}[htbp]
\epsscale{1.1}
\plotone{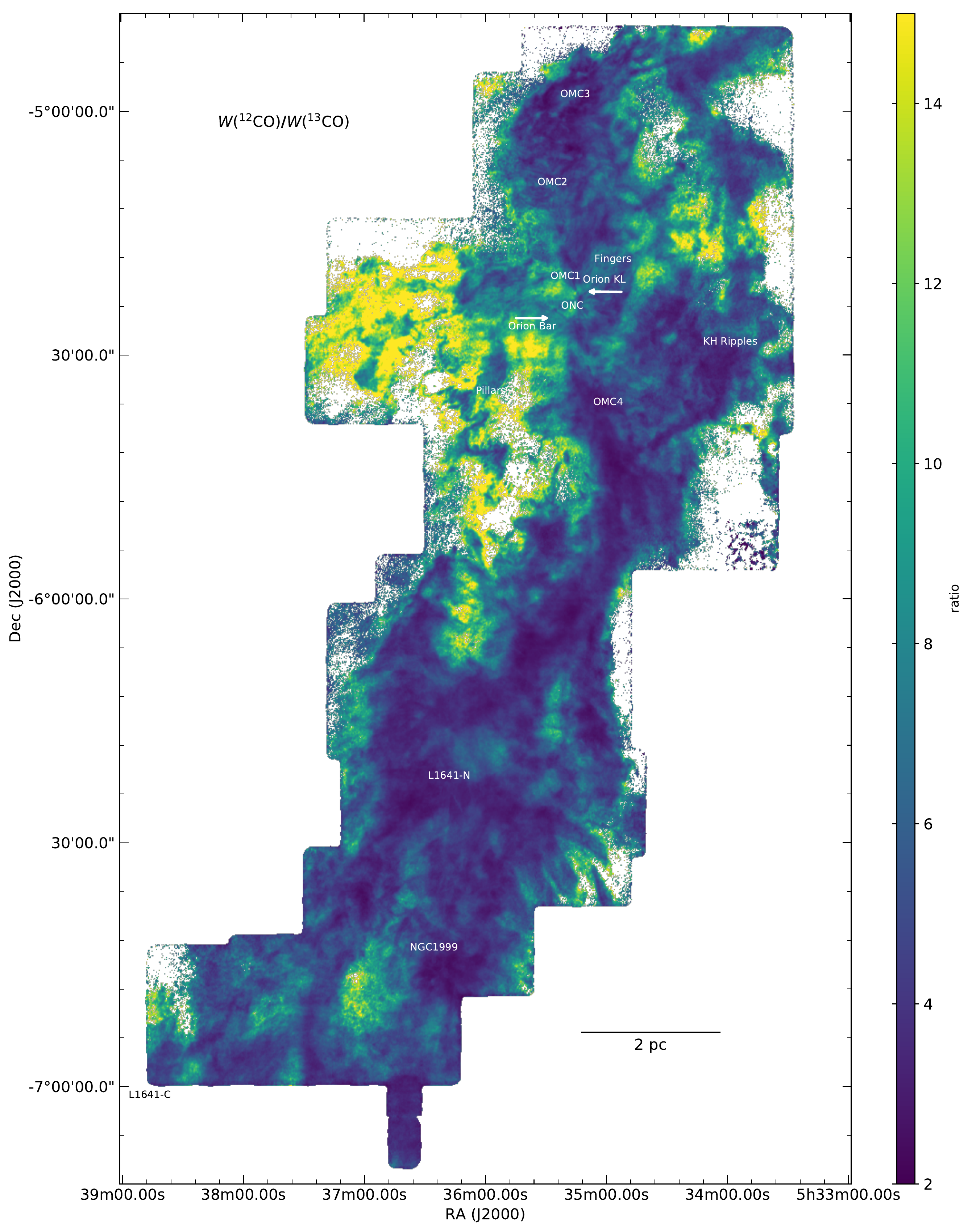}
\caption{Map of the ratio of $^{12}$CO(1-0) to $^{13}$CO(1-0) 
integrated intensities in the Orion A cloud. Masked areas (white) indicate  pixels
in the maps of Figures \ref{fig:12mom0} or \ref{fig:13mom0}
where SNR  $<$ 5.
\label{fig:ratio1213}}
\end{figure*}

In our maps, $^{12}$CO(1-0) is usually optically
thick and thus more sensitive to the line width 
and turbulence in the observed gas. 
By contrast, $^{13}$CO(1-0) is in general
optically thin (section \ref{subsec:coldens}) and 
more sensitive to column density variations 
\citep{2008ApJ...679..481P}. The $^{12}$CO to $^{13}$CO
ratio can therefore provide indications of the physical 
state of the molecular gas. Figure \ref{fig:ratio1213} 
shows the variation in the ratio of $^{12}$CO(1-0) 
integrated intensity, hereafter $W\rm (^{12}CO)$, 
to that of $^{13}$CO(1-0), hereafter $W\rm (^{13}CO)$,
across the mapped region of the Orion A cloud.
Pixels with SNR $<$ 5 in the 
integrated intensity maps of Figures \ref{fig:12mom0}
and \ref{fig:13mom0} are masked (white). 
The same integration range, (2.5 to 15 km s$^{-1}$) 
was used for both the $^{12}$CO and $^{13}$CO cubes.
Factors contributing less dramatically
to $W\rm (^{12}CO)$ and $W\rm (^{13}CO)$
include the CO abundance and the gas excitation. 
Excitation is regulated by gas volume density so that
low densities can lead to sub-thermal excitation, while at 
low temperatures, CO depletion can reduce the gas-phase 
abundance and resulting emission intensity. 
In relatively unshielded regions dissociation by 
FUV radiation can also reduce the abundance.

In Figure \ref{fig:ratio1213}, values of the 
$W\rm (^{12}CO)$/$W\rm (^{13}CO)$ ratio are $<$ 6 across 
most of the cloud, including the major ISF filament.
In some regions the ratio can be as low as
$\sim$ 2 (e.g., OMC-3, OMC-4, L1641-N, NGC 1999).
A comparison with Figures \ref{fig:13mom0} and
\ref{fig:coldens13} shows strong $^{13}$CO 
emission consistent with high column densities
in these same regions. Overall the trends are 
qualitatively similar to those noted by
\citet{2008ApJ...679..481P} in Perseus.

By contrast, much higher ratios ($\sim 15$) 
are found 
in the PDR regions to the east of ONC and 
OMC-4 where the photo-eroded pillars are located. 
Other high ratio regions include the area north-west
of the ``Fingers'' and south-east of NGC 1999. 
These are probably due to 
the combination of radiative transfer, selective 
photo-dissociation, and chemical fractionation
that is also seen in the Orion B cloud \citep{2017A&A...599A..98P}.
A transition border ($\sim$ 10) from low ratios
to high ratios can be seen surrounding the high column
density gas, indicative of selective photodissociation
\citep[e.g.,][]{2014A&A...564A..68S,2015ApJ...805...58K}.
A full understanding of
gas excitation and dissociation at both large
and small scales will require 
synthetic observations based on simulations 
that include radiative feedback and chemistry.

\section{Summary}

In this paper, we present the results 
obtained by combining new 
CARMA $^{12}$CO(1-0), $^{13}$CO(1-0), 
and C$^{18}$O(1-0) molecular line 
observations of an extended area of the Orion A cloud with 
complementary NRO45 measurements of the same region.
The footprint spans about 2 degrees in declination
and reaches from OMC-3 and OMC-2 regions 
in the north to the NGC 1999
and L1641-C complex in the south.
In particular, we provide practical details 
of our data combination methods, 
and describe our checks on the 
effectiveness and accuracy of our processes.  
We find that over a carefully defined range, 
the observed CARMA and NRO45 fluxes can be 
matched using a scale factor of 1.6. 
Likewise, a $\Delta$-variance analysis on the product images
show no artifacts due to the data combination. 

The breadth of the information on star formation
phenomena and processes that can be derived from the
combined data sets is very obvious in the images shown.
Integrated intensity maps of 
$^{12}$CO(1-0), $^{13}$CO(1-0), and C$^{18}$O(1-0) 
emission display a range of structures typical 
of every stage of early stellar evolution. 
Channel maps and images of velocity variations across the cloud  
indicate an overall large-scale velocity gradient 
from north to south. In addition, the different 
structures identified in the molecular line 
intensity maps often have particular kinematic
signatures. The potential scope and detail for 
kinematic studies is illustrated by position-velocity
diagrams of the Orion ISF feature. Maps of 
excitation temperature and column density trace
physical properties of the Orion A cloud as a 
whole and of individual features by themselves, 
and in the context of the larger cloud environment. 

Perhaps most importantly, by successfully 
combining high resolution interferometer maps
with lower resolution single-dish observations
of the Orion A cloud we have produced molecular line 
images of an extended star-forming cloud
with unprecedented detail. 
Studies of filaments, feedback processes, and cloud-scale star 
formation laws are already underway using these maps.
There is a wealth of information to be tapped
from these data. Maps presented here are currently  available 
\dataset[https://dataverse.harvard.edu/dataverse/CARMA-NRO-Orion]{https://dataverse.harvard.edu/dataverse/CARMA-NRO-Orion}
and  we anticipate making all three combined Orion A data cubes 
publicly available within 18 months.

\software{Astropy \citep{Astropy-Collaboration13}, Numpy \citep{numpy}, APLpy \citep{Robitaille12}, Matplotlib \citep{matplotlib}}

\facilities{CARMA, No:45m}

\acknowledgments 
We thank the anonymous referee for a thorough check on
the paper and helpful comments. We thank Amelia Stutz
for fruitful discussions.
This research was supported by the National Science 
Foundation, award AST-1140063, which 
also provided partial support for CARMA operations. 
CARMA operations were also supported by
the California Institute 
of Technology, the University of
California-Berkeley, the University of Illinois at
Urbana-Champaign, the University of Maryland College Park, and the
University of Chicago. The Nobeyama 45 m telescope 
is operated by the Nobeyama Radio 
Observatory, a branch of the National Astronomical Observatory 
of Japan. We are grateful to the staff at both 
NRO and CARMA for all their contributions. 
We thank the Yale Center for Research Computing for 
guidance and use of the research computing infrastructure.
We thank Rob Gutermuth for providing the Spitzer images.
HGA and JRF were partially funded by NSF award AST-1311825 while conducting this study.
JEP acknowledges the financial support of the European Research Council (ERC; project PALs 320620).
RJS acknowledges an STFC Ernest Rutherford fellowship.
V.O. was supported by the Deutsche
Forschungsgemeinschaft, DFG, through project number 177/2-2, 
and central funds of the DFG-priority program 1573 (ISM-SPP).
STS and ASM acknowledge funding by the Deutsche Forschungsgemeinschaft (DFG) via the Sonderforschungs\-bereich SFB 956 Conditions and Impact of Star Formation (subproject A4 and A6) and the Bonn-Cologne Graduate School.
RSK acknowledges financial support from the Deutsche Forschungsgemeinschaft via SFB 881, ``The Milky Way System'' (sub-projects B1, B2 and B8) and SPP 1573, ``Physics of the Interstellar Medium''. He also thanks for funding from the European Research Council via the ERC Advanced Grant ``STARLIGHT: Formation of the First Stars'' (project number 339177).
This research was carried out in part at the Jet Propulsion Laboratory, operated for NASA by the California Institute of Technology.

\bibliographystyle{aasjournal}
\bibliography{ref}

\appendix

\section{Channel maps.}\label{sec:chanmaps}

\begin{figure*}[htbp]
\epsscale{1.}
\plotone{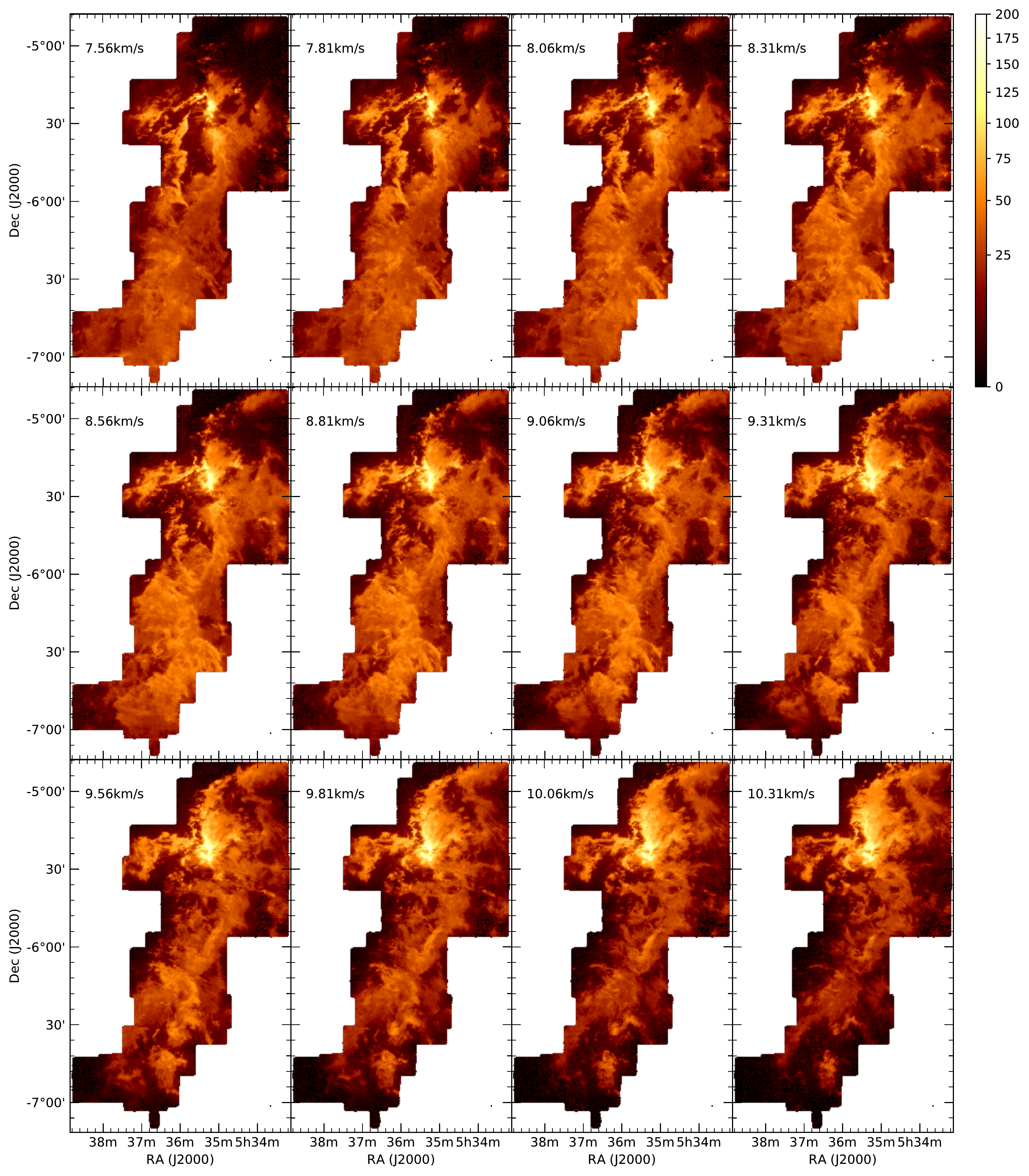}
\caption{
$^{12}$CO(1-0) channel maps from $V_{lsr}$ = 7.56 to 10.31 $\rm km~s^{-1}$.
The complete figure set (6 images) is available 
in the online journal.
\label{fig:chan12}}
\end{figure*}

\figsetstart
\figsetnum{\ref{fig:chan12}}
\figsettitle{$^{12}$CO(1-0) channel maps.}
 
\figsetgrpstart
\figsetgrpnum{\ref{fig:chan12}.1}
\figsetgrptitle{Channel-map figure 1.}
\figsetplot{chan12co9.pdf}
\figsetgrpnote{Channel maps from 0.06 to 2.81 $\rm km~s^{-1}$.}
\figsetgrpend
 
\figsetgrpstart
\figsetgrpnum{\ref{fig:chan12}.2}
\figsetgrptitle{Channel-map figure 2.}
\figsetplot{chan12co21.pdf}
\figsetgrpnote{Channel maps from 3.06 to 5.81 $\rm km~s^{-1}$.}
\figsetgrpend
 
\figsetgrpstart
\figsetgrpnum{\ref{fig:chan12}.3}
\figsetgrptitle{Channel-map figure 3.}
\figsetplot{chan12co33.pdf}
\figsetgrpnote{Channel maps from 6.06 to 8.81 $\rm km~s^{-1}$.}
\figsetgrpend
 
\figsetgrpstart
\figsetgrpnum{\ref{fig:chan12}.4}
\figsetgrptitle{Channel-map figure 4.}
\figsetplot{chan12co45.pdf}
\figsetgrpnote{Channel maps from 9.06 to 11.81 $\rm km~s^{-1}$.}
\figsetgrpend
 
\figsetgrpstart
\figsetgrpnum{\ref{fig:chan12}.5}
\figsetgrptitle{Channel-map figure 5.}
\figsetplot{chan12co57.pdf}
\figsetgrpnote{Channel maps from 12.06 to 14.81 $\rm km~s^{-1}$.}
\figsetgrpend
 
\figsetgrpstart
\figsetgrpnum{\ref{fig:chan12}.6}
\figsetgrptitle{Channel-map figure 6.}
\figsetplot{chan12co69.pdf}
\figsetgrpnote{Channel maps from 15.06 to 17.81 $\rm km~s^{-1}$.}
\figsetgrpend
 
\figsetend

\newpage

\begin{figure*}[htbp]
\epsscale{1.}
\plotone{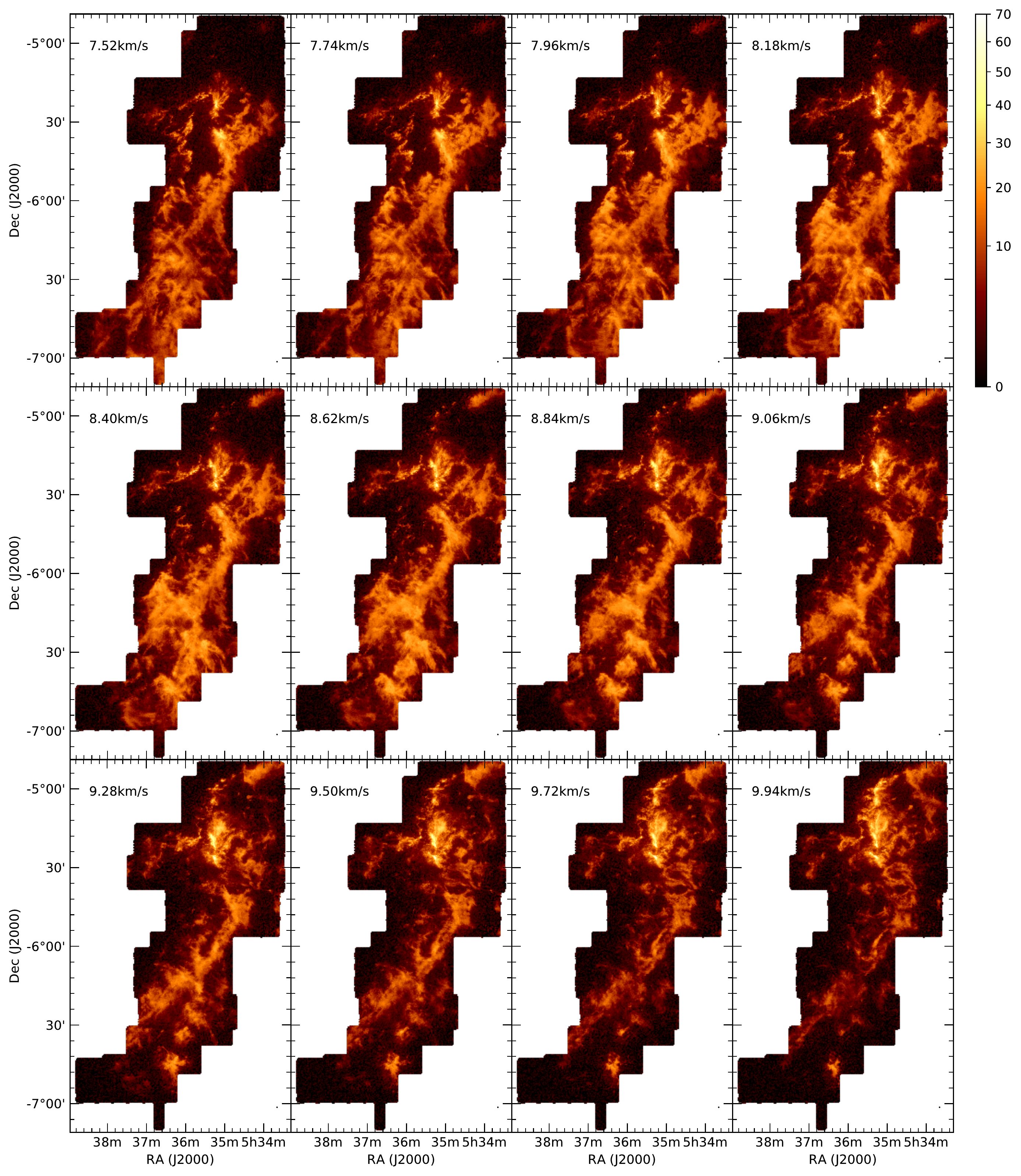}
\caption{
$^{13}$CO(1-0) channel maps from $V_{lsr}$ = 7.52 to 9.94 $\rm km~s^{-1}$.
The complete figure set (6 images) is available 
in the online journal.
\label{fig:chan13}}
\end{figure*}

\figsetstart
\figsetnum{\ref{fig:chan13}}
\figsettitle{$^{13}$CO(1-0) channel maps.}
 
\figsetgrpstart
\figsetgrpnum{\ref{fig:chan13}.1}
\figsetgrptitle{Channel-map figure 1.}
\figsetplot{chan13co0.pdf}
\figsetgrpnote{Channel maps from 0.04 to 2.46 $\rm km~s^{-1}$.}
\figsetgrpend
 
\figsetgrpstart
\figsetgrpnum{\ref{fig:chan13}.2}
\figsetgrptitle{Channel-map figure 2.}
\figsetplot{chan13co12.pdf}
\figsetgrpnote{Channel maps from 2.68 to 5.10 $\rm km~s^{-1}$.}
\figsetgrpend
 
\figsetgrpstart
\figsetgrpnum{\ref{fig:chan13}.3}
\figsetgrptitle{Channel-map figure 3.}
\figsetplot{chan13co24.pdf}
\figsetgrpnote{Channel maps from 5.32 to 7.74 $\rm km~s^{-1}$.}
\figsetgrpend
 
\figsetgrpstart
\figsetgrpnum{\ref{fig:chan13}.4}
\figsetgrptitle{Channel-map figure 4.}
\figsetplot{chan13co36.pdf}
\figsetgrpnote{Channel maps from 7.96 to 10.38 $\rm km~s^{-1}$.}
\figsetgrpend
 
\figsetgrpstart
\figsetgrpnum{\ref{fig:chan13}.5}
\figsetgrptitle{Channel-map figure 5.}
\figsetplot{chan13co48.pdf}
\figsetgrpnote{Channel maps from 10.60 to 13.02 $\rm km~s^{-1}$.}
\figsetgrpend
 
\figsetgrpstart
\figsetgrpnum{\ref{fig:chan13}.6}
\figsetgrptitle{Channel-map figure 6.}
\figsetplot{chan13co60.pdf}
\figsetgrpnote{Channel maps from 13.24 to 15.66 $\rm km~s^{-1}$.}
\figsetgrpend
 
\figsetend

\newpage

\begin{figure*}[htbp]
\epsscale{1.}
\plotone{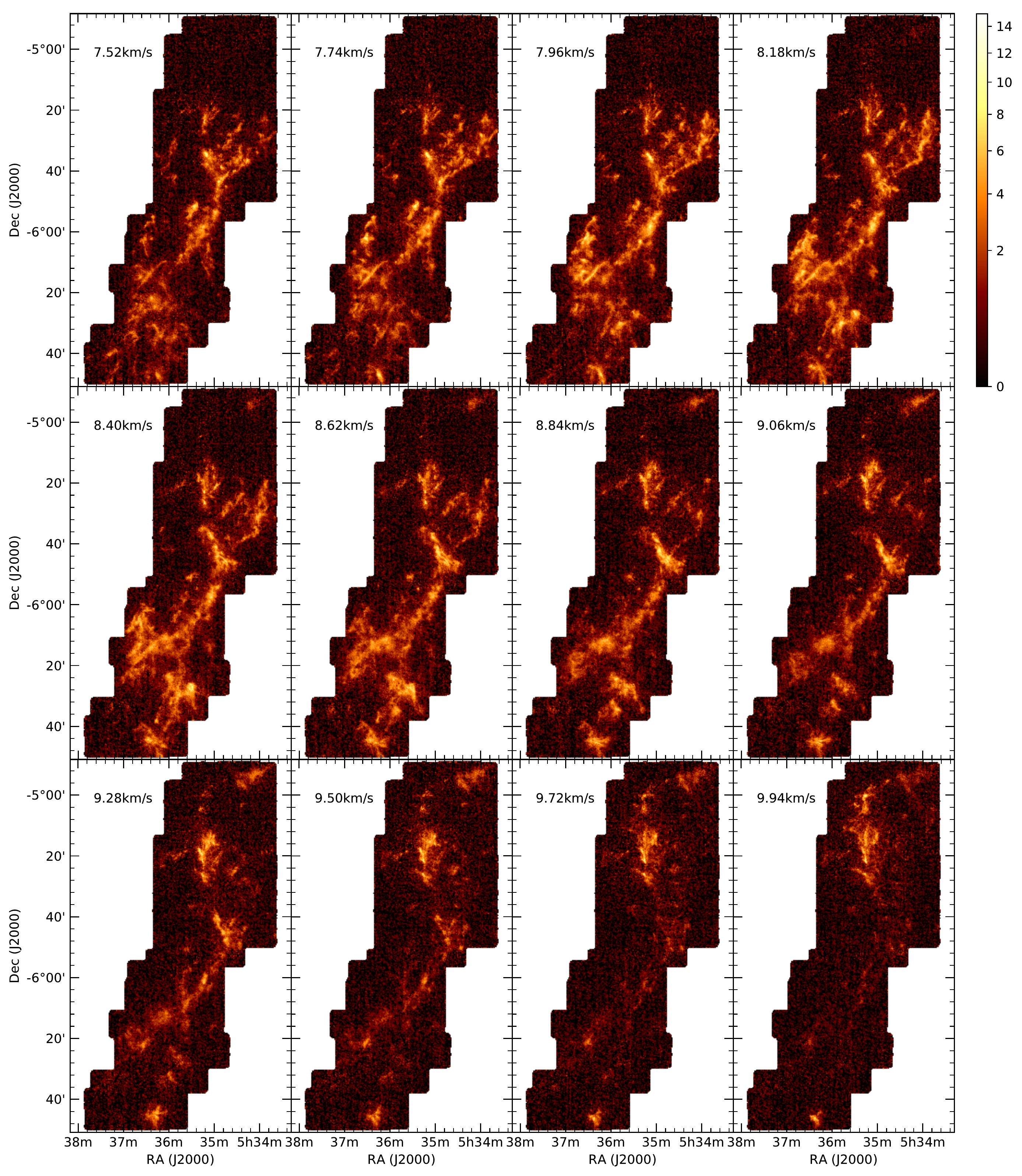}
\caption{
C$^{18}$O(1-0) channel maps from $V_{lsr}$ = 7.52 to 9.94 $\rm km~s^{-1}$.
The complete figure set (6 images) is available 
in the online journal.
\label{fig:chan18}}
\end{figure*}

\figsetstart
\figsetnum{\ref{fig:chan18}}
\figsettitle{C$^{18}$O(1-0) channel maps.}
 
\figsetgrpstart
\figsetgrpnum{\ref{fig:chan18}.1}
\figsetgrptitle{Channel-map figure 1.}
\figsetplot{chanc18o0.pdf}
\figsetgrpnote{Channel maps from 0.04 to 2.46 $\rm km~s^{-1}$.}
\figsetgrpend
 
\figsetgrpstart
\figsetgrpnum{\ref{fig:chan18}.2}
\figsetgrptitle{Channel-map figure 2.}
\figsetplot{chanc18o12.pdf}
\figsetgrpnote{Channel maps from 2.68 to 5.10 $\rm km~s^{-1}$.}
\figsetgrpend
 
\figsetgrpstart
\figsetgrpnum{\ref{fig:chan18}.3}
\figsetgrptitle{Channel-map figure 3.}
\figsetplot{chanc18o24.pdf}
\figsetgrpnote{Channel maps from 5.32 to 7.74 $\rm km~s^{-1}$.}
\figsetgrpend
 
\figsetgrpstart
\figsetgrpnum{\ref{fig:chan18}.4}
\figsetgrptitle{Channel-map figure 4.}
\figsetplot{chanc18o36.pdf}
\figsetgrpnote{Channel maps from 7.96 to 10.38 $\rm km~s^{-1}$.}
\figsetgrpend
 
\figsetgrpstart
\figsetgrpnum{\ref{fig:chan18}.5}
\figsetgrptitle{Channel-map figure 5.}
\figsetplot{chanc18o48.pdf}
\figsetgrpnote{Channel maps from 10.60 to 13.02 $\rm km~s^{-1}$.}
\figsetgrpend
 
\figsetgrpstart
\figsetgrpnum{\ref{fig:chan18}.6}
\figsetgrptitle{Channel-map figure 6.}
\figsetplot{chanc18o60.pdf}
\figsetgrpnote{Channel maps from 13.24 to 15.66 $\rm km~s^{-1}$.}
\figsetgrpend
 
\figsetend

\end{document}